\documentclass[manuscript,screen,authorversion,nonacm]{acmart}
\setcopyright{acmlicensed}
\acmJournal{PACMHCI}
\acmYear{2023} \acmVolume{7} \acmNumber{CSCW1} \acmArticle{115} \acmMonth{4} \acmPrice{15.00}\acmDOI{10.1145/3579591}

\usepackage{color, soul}
\usepackage{tabularx}
\usepackage{graphicx}
\usepackage{booktabs} 
\usepackage{multirow}
\usepackage{color}
\usepackage{amsmath}
\usepackage{latexsym}
\usepackage{geometry}
\usepackage[english]{babel}
\usepackage{blindtext}
\usepackage{graphicx}
\usepackage{subfigure}
\usepackage{indentfirst}
\usepackage{ulem}
\usepackage{float}
\usepackage{indentfirst}
\usepackage{setspace}
\usepackage{makecell}
\usepackage{pifont}
\usepackage{hyperref}
\hypersetup{
    colorlinks=true,
    linkcolor=blue,
    filecolor=blue,      
    urlcolor=blue,
    citecolor=cyan,
}
\usepackage{hyperref} 
\usepackage{fontawesome} 

\newcommand{\sy}[1]{\textcolor{black}{#1}}

\newcommand{\cltext}[1]{\textcolor{black}{#1}}
\newcommand{\newsy}[1]{\textcolor{black}{#1}}

\AtBeginDocument{%
  \providecommand\BibTeX{{%
    \normalfont B\kern-0.5em{\scshape i\kern-0.25em b}\kern-0.8em\TeX}}}

\begin{document}

\title[Wizundry: A Cooperative Wizard of Oz Platform]{Wizundry: A Cooperative Wizard of Oz Platform for Simulating Future Speech-based Interfaces with Multiple Wizards}


\author{Siying Hu}
\affiliation{%
    \department{School of Creative Media and Department of Computer Science}
    \institution{City University of Hong Kong}
  \city{Hong Kong SAR}
  \country{China}}
\email{siyinghu-c@my.cityu.edu.hk}

\author{Hen Chen Yen}
\affiliation{%
    \department{Department of Computer Science}
    \institution{City University of Hong Kong}
  \city{Hong Kong SAR}
  \country{China}}
\affiliation{%
    \department{David R. Cheriton School of Computer Science}
    \institution{University of Waterloo}
  \city{Waterloo, Ontario}
  \country{Canada}}
\email{r4yen@uwaterloo.ca}

\author{Ziwei Yu}
\affiliation{%
    \department{School of Creative Media}
    \institution{City University of Hong Kong}
  \city{Hong Kong SAR}
  \country{China}}
\affiliation{%
    \institution{Honor Device Co Ltd}
  \city{Shenzhen}
  \country{China}}
\email{yzw2109@gmail.com}

\author{Mingjian Zhao}
\affiliation{%
    \department{Department of Computer Science}
    \institution{City University of Hong Kong}
  \city{Hong Kong SAR}
  \country{China}}
\affiliation{%
    \department{Department of Computer Science}
    \institution{Washington University in St. Louis}
  \city{St. Louiso}
  \country{U.S}}
\email{mingjian.z@wustl.edu}\textbf{}

\author{Katie Seaborn}
\affiliation{%
  \department{Department of Industrial Engineering and Economics}
  \institution{Tokyo Institute of Technology}
  \city{Tokyo}
  \country{Japan}}
\email{seaborn.k.aa@m.titech.ac.jp}

\author{Can Liu}
\authornote{Corresponding Author}
\affiliation{%
    \department{School of Creative Media}
    \institution{City University of Hong Kong}
  \city{Hong Kong}
  \country{China}}
\email{canliu@cityu.edu.hk}

\renewcommand{\shortauthors}{Siying Hu et al.}

\begin{abstract}
Wizard of Oz (WoZ) as a prototyping method has been used to simulate intelligent user interfaces, particularly for speech-based systems. However, as our societies' expectations on artificial intelligence (AI) grows, the question remains whether a single Wizard is sufficient for it to simulate smarter systems and more complex interactions. Optimistic visions of 'what artificial intelligence (AI) can do' places demands on WoZ platforms to simulate smarter systems and more complex interactions. This raises the question of whether the typical approach of employing a single Wizard is sufficient. Moreover, while existing work has employed multiple Wizards in WoZ studies, a multi-Wizard approach has not been systematically studied in terms of feasibility, effectiveness, and challenges. We offer Wizundry, a real-time, web-based WoZ platform that allows multiple Wizards to collaboratively operate a speech-to-text based system remotely. We outline the design and technical specifications of our open-source platform, which we iterated over two design phases. We report on two studies in which participant-Wizards were tasked with negotiating how to cooperatively simulate an interface that can handle natural speech for dictation and text editing as well as other intelligent text processing tasks. We offer qualitative findings on the Multi-Wizard experience for Dyads and Triads of Wizards. Our findings reveal the promises and challenges of the multi-Wizard approach and open up new research questions. 

\end{abstract}




\maketitle

\section*{Citation}

\noindent
Siying Hu, Hen Chen Yen, Ziwei Yu, Mingjian Zhao, Katie Seaborn, and Can Liu. 2023. Wizundry: A Cooperative Wizard of Oz Platform for Simulating Future Speech-based Interfaces with Multiple Wizards. \textit{Proc. ACM Hum.- Comput. Interact. 7}, CSCW1, Article 115 (April 2023), 34 pages. \url{https://doi.org/10.1145/3579591} \\

\noindent
The final publication is available via ACM at \url{https://dl.acm.org/doi/10.1145/3579591}.

\section{Introduction}

Speaking is considered to be one of the most natural input modalities for text input and for general interactions with a computer. Current statistics show that voice search is used by 41\% of adults at least once a day \footnote{https://www.dbswebsite.com/blog/trends-in-voice-search/}, showing how the use of speech as primary modality for user engagement is becoming increasingly widespread \cite{10.1145/2163.358100}. 
Smart speakers, conversational agents (CAs), dictation interfaces, and other speech-based systems and environments can be found in our homes, workplaces, public spaces, and even in our pockets. Advanced speech recognition has become an integral part of the operating systems of modern smartphones for supporting speech-based text input, such as in the Google Pixel 6.
This trend is further supported by forecasts expecting voice commerce to become an US\$80 billion business by 2023 \footnote{https://techcrunch.com/2019/02/12/report-voice-assistants-in-use-to-triple-to-8-billion-by-2023}. 

Most speech-based interfaces rely on advancements in machine learning (ML) technologies. Natural language processing (NLP), in particular, provides the basis for understanding voice commands and speech patterns from human users and assists in generating appropriate and engaging vocal responses from the system itself \cite{chowdhury2003natural,hirschberg2015advances}. Text-to-speech (TTS) then gives rise to the voice of the machine: text generated from NLP algorithms is transformed into an audible output using voice samples and/or synthesizers \cite{dutoit1997introduction}. Both of these core components of speech-based systems are complex and may be costly when developing specific applications, tasks, and activities and/or when it comes to employing custom voices\footnote{https://venturebeat.com/2021/01/15/amazon-launches-alexa-custom-assistant-to-let-brands-build-their-own-voice-assistants/}. Therefore, to reduce the risk of future costs and delays, researchers and designers of such systems use Wizard of Oz (WoZ) method to study the user experience of a future system that is simulated by a human ``Wizard.'' In the early 1980s, Bell Labs  developed the first example of a WoZ system: a ``listening typewriter'' \cite{10.1145/2163.358100} and since then, WoZ has been used in an ever-increasing number of HCI studies \cite{10.1145/3432942,DAHLBACK1993258} with more than one quarter of studies involving speech-based interaction specifically \cite{10.1145/3386867,10.1145/3405755.3406148}. In a large-scale survey of the literature covering two decades of research work, Seaborn et al. \cite{10.1145/3386867} found that 27\% of all studies on voice-based agents, interfaces, and environments relied on WoZ for controlling the system and/or performing the voice in real-time. 

\sy{However, while WoZ is efficient in terms of time and resources it requires, it has its drawbacks and challenges. Most WoZ platforms employ a single Wizard to perform the simulation, who is then required to react quickly and consistently to meet the desired standards in view of task performance, response time, error rates, and so on \cite{schlogl2013managing,salber1993applying}. This places both predictable and unpredictable demands on the Wizard, as he/she needs to juggle multiple variables while dynamically adjusting the script to accommodate unexpected user reactions and technical challenges. Moreover, despite this already high cognitive overload,  technological advancements have increased the expected performance of systems, and therefore Wizards. In particular, speech-based systems are expected to move towards greater complexity in response to the heightened demand for dynamic and open-ended interactions, potentially with non-voice complements \cite{budzianowski2018multiwoz,10.1145/3411764.3445536}. 
This added demand for multi-modality makes the Wizard's task and the information bandwidth more complex \cite{salber1993applying, de1998visual}, yet he/she cannot fumble because if they are  too slow to respond, the end-user may react negatively, avoid using the new features being tested, or make mistakes that would not otherwise occur in a real system. Therefore, when the task is complex, ill-defined, multi-modal or multi-threaded, and/or dynamic, the burden on the Wizard can increase dramatically, to the point that they can become "lost in space," i.e., unable to determine where they need to go in the Wizarding interface \cite{10.1007/978-3-540-73105-4_26, 8673205}. This raises an important question: How can we ensure that WoZ remains an effective strategy when designing near-future, innovative voice-based systems?}

\sy{One solution is to bring on multiple Wizards who can combine their cognitive resources and work collaboratively to simulate the system: the \emph{Multi-Wizard approach}. 
For example, multiple Wizards have been employed to} simulate multi-modal interfaces, where each Wizard is responsible for a different user input channel \cite{coutaz1996neimo,Salber93requirementsfor}. 
\sy{
Nevertheless, problems can occur when the Wizards need to cooperatively sync their responses, share information across channels, or fuse modalities in some way \cite{salber1993applying}.
Non-Wizard researchers may also be working behind the curtains to aid in the simulation \cite{porcheron2021pulling,martelaro2020using}. 
As such, most Multi-Wizard configurations in the multi-modal space aim to alleviate cognitive overload and guarantee consistent behaviour across Wizards and supporting team members \cite{salber1993applying,DAHLBACK1993258}.}
\sy{Even so, no work to date has explored the Wizarding experience, i.e., Wizard-as-user studies. Are Multi-Wizard simulations of advanced systems feasible? How can we best support this collaborative task? Additionally, multiple Wizards may be leveraged for purposes other than supporting multi-model WoZ setups. In particular, a Multi-Wizard approach could be valuable for simulating advanced voice- and/or speech-based systems.}
\sy{
Engineers and designers are now exploring more complex systems such as: innovative note-taking tools for meetings and conferences; advanced editing features in dictation systems and multi-agent contexts; "smart" conversations involving, for instance, time-sensitive decision-making drawn from vast knowledge bases, and more multitask unified model algorithm. 
These near-future offerings 
involve more advanced tasks on the system side. Compared to simulating NLP or TTS, simulating these envisioned interfaces is likely to require a greater number and varieties of actions beyond specific modalities and thus place more demand on Wizards working alone and cooperatively. 
Researchers can also study the relative effects of advanced voice and/or speech contexts on Wizard performance and cognitive load, as well as how to provide optimal end-user experience in these near-future voice-based contexts. Furthermore, in the spirit of Open Science \cite{spellman2017open}, an open-source tool would be ideal for transparency, allowing extensions for different contexts and research questions, and enabling future generalizability across studies. As yet, no such tool exists.}

To this end, we offer \textit{Wizundry}, a new web-based WoZ platform for enabling multiple Wizards to simulate speech-based interfaces remotely and in real-time. Wizundry provides a modular and configurable interface that builds on two core functionalities: Speech-to-Text (STT) and Text-to-Speech (TTS), which can be used together or separately. STT is used in interface components for Wizards to edit, segment, tag and highlight the transcribed text, and to simulate intelligent text processing features. TTS is used to enable Wizards to create, edit and play speech messages or responses to ``end-users''. As a collaboration tool, it provides transparency and visibility of each Wizards' actions, and allow the Wizards to set their own strategies for cooperation and division of labor. Using an iterative design process, we created two versions of Wizundry and conducted two user studies to evaluate the feasibility and effectiveness of this approach as well as the design of our platform. Our studies identified challenges that hamper the Wizards' cooperation, as well as the strengths and weaknesses of a variety of strategies explored by the participants who took on the Wizard's role. We also found that our platform was able to support fast iterations of the intelligent interface design based on STT and/or TTS.

Our contributions are threefold. First, the design and technical specifications of Wizundry, which includes our open-source code repository\footnote{https://github.com/erfilab/Wizundry-MultiWizard}. Second, findings from two studies on \sy{Single-Wizardry} and Multi-Wizardry situations, including cooperation strategies for designing Multi-Wizard WoZ systems for speech-based interfaces and potentially for other intelligent systems. Thirdly, since this work acts as a timely and much needed addition to the growing areas of WoZ prototyping for speech-based interfaces in contemporary and future applications, our findings shed light on underdeveloped areas in WoZ methodology and open up new research questions.

\section{Related Work}

We will first review well-known Wizard of Oz tools and platforms, then we will discuss examples that are specific to voice interaction. Finally, we situate our work within the Multi-Wizard domain, with focus on methods for evaluating the Wizard UX.

\subsection{Wizard of Oz Tools and Platforms}
Wizard of Oz is a standard approach to evaluating prototypes at various stages of development and at varying levels of fidelity \cite{DAHLBACK1993258}. Often the most difficult, expensive, or impossible to develop aspects of the system are left up to WoZ for early testing. In one of the earliest examples, IBM created a WoZ platform for a user interface in their personal computer, which was still an uncharted territory in the mid-1980s \cite{doi:10.1177/154193128502900515}. Mander et al. \cite{10.1145/169059.169215} created Turvy, a WoZ agent for exploring futuristic human-agent interactions that were difficult or impossible to achieve in the 1990s. Users were asked to ``teach'' the agent by talking to it while pointing at objects, and the Wizard behind the curtain registered and responded to this information, allowing for smooth and prescient testing of this futuristic vision of technology. Many WoZ platforms are essentially early versions of the final products, and so do not 
have one-size-fit-all WoZ platform that can cater to all variations.

Within research spaces of HRI, the end goal may not be a new platform, but rather a new understanding of how people use and relate to computers. Platforms that allow for WoZ have thus become widespread within research, such as Aldebaran-SoftBank's humanoid robot Nao\footnote{ https://www.softbankrobotics.com/emea/en/nao} or the general platform ZBOS applied to a care robot called Zora 
\footnote{https://zorarobotics.be/zbos-zora}. In particular, it comes with a simple programmable interface, text-to-speech (TTS) tool, and 3D simulation environment that have been frequently co-opted for WoZ in research (e.g., \cite{10.1145/3386867,10.5898/JHRI.1.1.Riek,sandygulova2018age,7745134}).

Platform specificity (or ``one-off'' prototypes, such as Turvy) and researcher flexibility (as with the widespread uptake of the Nao) are common orientations to WoZ in research. However, there are some general WoZ tools and platforms that can be integrated into a system under development or customized for the research question under study. As Cambre and Kulkarni recently identified \cite{10.1145/3405755.3406148}, these prototyping methods may be designed to allow for WoZ of a specific feature or function (e.g., elicitation methods, dialogue management systems) and not others. Many of these have been developed for or as a result of research. SUEDE~\cite{10.1145/354401.354406} was a visual prototyping tool designed to help researchers quickly create speech UIs without needing to do much coding. WebWOZ \cite{10.1145/1822018.1822035} was one of the first web-based, cross-platform WoZ tools for research involving dialogues between people and/or computers. Marionette \cite{10.1145/2909824.3020256} was developed to fill the gap in WoZ platforms for human-vehicle interaction, an emerging area of research facilitated by the increased use of voice-based, hands-free smart agents in vehicles (or as the vehicle itself). WebApp \cite{10.1145/3343413.3377941,schlogl2013webwoz} was designed to address the gap in text-based conversational interface WoZ tools for search, particularly collaborative search within Slack\footnote{https://slack.com/intl/zh-hk/}. Commercial platforms, such as Voiceflow\footnote{See https://www.voiceflow.com/} and Alexa Skills Kit\footnote{See https://developer.amazon.com/en-US/alexa/alexa-skills-kit} can also be adapted for research as well as other purposes, but may be costly, limited to certain technologies (e.g., Amazon Alexa), or otherwise restricted due to proprietary matters.

Most WoZ platforms for voice interaction, both classic and modern, have focused on the problem of speech recognition and speech output in information retrieval contexts: natural language processing (NLP) \cite{diaper1989wizard,richards1984should,guindon1987grammatical,jonsson1988talking,whittaker1989user}. A survey of HRI works revealed that 72.2\% of WoZ approaches to robots were selected for this reason \cite{10.5898/JHRI.1.1.Riek}. Shamekhi et al. \cite{shamekhi2018face} compared voice-only and bodied versions of conversational agent as a collaborator in a group-based decision-making activity wherein WoZ was used to detect speakers' intent and provide a response. As voice interfaces have proliferated and matured, new factors and concerns have arisen. Children, for instance, have been identified as a new and potentially vulnerable user group. Yarosh et al. \cite{10.1145/3202185.3202207} created three WoZ speech interfaces to study the patterns of questioning and answering amongst children. Context also matters as research moves ``into the wild'' in field studies. Braun et al. \cite{10.1145/3290605.3300270} developed a multi-personality voice-based smart vehicle agent that was tested under realistic driving conditions. The Wizard was responsible for changing the response patterns of the agent at opportune moments.

As this body of works show, speech-based interface is a booming area of study for which robust and ideally general WoZ platforms are necessary. We contribute to it by extending the remote, web-based, open source, cross-platform, and multi-modal approaches explored so far to speech-based interaction and multiple Wizards. Next we turn to the challenge of supporting multiple Wizards in the handling of increasingly sophisticated interaction scenarios of the near future--by elaborating on the Wizards' own experiences as users of such technologies in a collaborative context.

\subsection{Towards Wizardry: Studies of the Wizard Experience}

Taking on the role of a Wizard can be a challenging task. Training is often required, but it may not eliminate or mitigate the difficulties the Wizard may experience. Difficulties are compounded when the Wizard does not have assistance and communicate directly with others involved in conducting the study, for fear of revealing the fiction of the Wizardry to the participant \cite{10.1145/3432942}. Little is known about the exact challenges that Wizards face, as the Wizard's experience is not often reported in the literature. Still, some challenges have been identified.

One challenge is the cognitive load. Shin, Oh, and Lee \cite{8673205} identified four areas in which a Wizard may experience cognitive overload when playing the role of a conversational robot as they need to multitask between paying attention, decision-making, task-execution, and reflection. Their solution was to introduce an autonomous agent collaborator to assist the Wizard in their Wizardry. Large et al. ~\cite{10.1145/3003715.3005408} tested a digital driving assistant (DDA) in a vehicle driving simulation environment. Their setup featured two Wizards, each responsible for one cognitively demanding task: conversing with the driver and assisting with search and navigation. This work reduced cognitive load by dividing labor.

Another challenge is multi-modality. Salber and Coultaz ~\cite{salber1993applying} developed a WoZ platform called Neimo to support multi-modal designs. They considered that each modality could be the responsibility of each individual Wizard, e.g., a Speech Wizard, Face Wizard, and Mouse Wizard. Cohen et al. ~\cite{10.1145/1452392.1452419} explored this notion explicitly in the design of a WoZ platform that allows Multi-Wizard control of the key interface components, pen and speech, dividing the labour between two Wizards roles: content and user pen output analysis.

A third challenge relates to NLP, which is the foremost reason for employing WoZ in voice-based systems, conversational agents, and speech interfaces because they entail open-ended responses. Yarosh et al. \cite{10.1145/3202185.3202207} provided Wizards with a list of preset responses but allowed for these to be changed on the fly. The necessity to allow for open-ended responses from the Wizard was also found by Vtyurina et al. \cite{10.1145/3027063.3053175}, who recognized that searching for preset passages could impede the performance of the Wizard compared to simply allowing them to write the passage off the cuff.

A related challenge comes from the fact that computers have yet to master the detection of implicit and/or nonverbal cues, whereas our Wizards certainly could. Vtyurina et al. \cite{10.1145/3173574.3173782} identified several ways in which users conveyed an expectation of a response from a conversational cooking assistant. For instance, when the user was ready to move on to the next step, they did not explicitly say so, but rather used a vague endorsements such as ``okay,'' which the Wizard could understand and act upon.

The literature so far has focused on how to support solo Wizardry. Multiple Wizards can be trained to carry out the work in shifts (e.g., \cite{10.1145/3202185.3202207}), but this can introduce inconsistent experiences for the participants. While it may be effective to apply a truly \emph{Multi-Wizard} approach in which multiple Wizards working in concert, this approach has been far less explored with only a few reported examples. For instance, a couple of studies \cite{1374824,marge2017applying} have considered dividing the labour involved in controlling a robot within a co-present WoZ context: one Wizard controlling the dialogue and the other controlling the robot's movement. Another work extended this idea to a remote context \cite{bonial2017laying}, however, none of them has explicitly evaluated the Wizards' experience. A few (e.g., \cite{serrano2010wizard}) have proposed Multi-Wizardy in future work but these ideas have yet to materialize. 

Whether considering solo- or Multi-Wizard contexts, evaluating the experience of the Wizard as a performer and/or user of the WoZ platform has received little attention in general. Riek \cite{10.5898/JHRI.1.1.Riek} found that most studies constrained what the Wizard could do (90.7\%) but not what they were able to perceive (11\%). Very few reported measuring Wizard error (3.7\%) or reported pre-experiment Wizard training (5.4\%). In these cases, the effects of these factors on the act of Wizardry as well as the experience of the Wizard are not known.

In this work, we focus on the experience of WoZ from the standpoint of the Wizard, treating the Wizard as an essential user of the platform and their experience as a key variable in the success of the WoZ approach for end-users. Additionally, we evaluate not just the Wizard's experience as a solo actor, but also their experiences within a collaborative Multi-Wizard context. To the best of our knowledge, this is the first work to do so, generally and within WoZ for speech interfaces. As our platform essentially supports intelligent features by leveraging input from multiple human Wizards , the design has similarities with some real-time crowd-sourcing platforms. Thus, we will now review related literatures on these plaforms to explicate the differences between our approaches.

\subsection{Real-time Crowd-sourcing Platforms}
Real-time crowd-sourcing platforms provide similar functionalities to our system, in terms of enabling the division of labor and in facilitating the workflow for combining their outputs. 
\sy{Researchers have made use of advanced interface designs (e.g., visualizations of collaborator actions and separate work-spaces) and task allocation mechanisms in workflows (e.g., complex-tasks-to-micro-tasks), enabling workers to collaborate freely and efficiently to complete the goal \cite{lasecki2015apparition, huang2017supporting}. Some have investigated synchronous crowd contexts wherein a large number of disparate users jointly manipulate a user interface) \cite{lasecki2011real}}, whereas others, such as Lasecki et al. \cite{lasecki2011real}, aggregated disparate user entries into a single application workflow so as to return work results quickly to crowd-workers. In other words, this approach combines multiple real-time contributions into a single command. Similarly, Kim et al. \cite{kim2013cobi} designed Cobi to reduce work conflicts by providing visual indicators of each worker's task status, to decrease the time consumed in resolving conflicts among workers. 
\sy{In line with this approach, Wizundry also provides visual indicators for each Wizard, allowing for action transparency for efficient collaboration, especially by providing real-time cues where synchronized tasks may be possible.} To address such challenges,
Bernstein et al. \cite{bernstein2010soylent} split complex writing and editing work into several micro-tasks, assigning various people to work synchronously but
independently by applying a sequential, `Find-Fix-Verify' method.
\sy{This
enabled pre-assignment of different tasks to each crowd-worker on separate working interfaces.
Inspired by these design considerations, we designed our toolkit with separate work interfaces that support the division of speech-based interface simulation tasks, while enabling us to investigate the division of labour that can emerge among multiple Wizards.
}

\sy{Who manages the workflow is another key consideration. Lasecki et al. \cite{lasecki2015apparition} considered this issue in the context of WoZ-driven systems, which are often one-offs that still require significant development work. They designed their system Apparition}
\sy{to provide a "lightweight write-locking mechanism" so that workers could self-manage task delegation in real-time while also communicating with each other about what components they are modifying and what tasks they are working on. This mechanism allowed workers to self-manage task delegation in real-time.
Similarly, we designed Wizundry to enable multiple independent Wizards to self-allocate simulation tasks, and to synchronize this information with other Wizards to improve the Multi-Wizard workflow that results in a better end-user experience.}

\sy{Following these previous studies on crowd interface design, especially in view of multi-worker workflows, collaborative mechanisms, and the collocation of tasks, we designed our platform with operational independence from the Wizard's role. As a research tool, the Wizundry intelligent user interface leans on these previous endeavours by supporting cooperation among multiple Wizards who may take on unpredictable, ill-defined, and shifting tasks in real-time.} We also aimed to reduce the challenges and workload experienced by Wizards by increasing their team capacity. In the next section, we describe the design of our Multi-Wizard platform: Wizundry.

\section{Designing and Testing Wizundry 1.0}
Wizundry 1.0 is a Multi-Wizard WoZ platform designed for prototyping dictation-based speech interfaces. 
\sy{We drew from existing designs and research on collaboration strategies in text-based crowdsourcing works \cite{lin2021sync, hartmann2010would, bernstein2010soylent} while building on the foundation of existing WoZ platforms \cite{10.1145/3386867,10.1145/169059.169215,10.1145/3405755.3406148,10.1145/3343413.3377941,schlogl2013webwoz}. We also extended this work by focusing on the Multi-Wizard context, especially how this context affects the Wizard's UX (Wizard-as-user) \cite{8673205,10.1145/1452392.1452419,10.1145/3027063.3053175}. On this point, we anchored our design work on supporting the division of tasks among Wizards and allowing self-organization for collaboration \cite{salber1993applying, DAHLBACK1993258}. In choosing the context of a speech input/output interface, we provided a WoZ tool that can simulate a range of emerging speech technologies, as well as offering an advanced intelligent system context of use for researchers to study advanced forms of WoZ, especially in Multi-Wizard collaboration scenarios. 
Our goal was to improve the end-user experience of WoZ by enhancing how Wizards work together. In Wizundry, Multiple Wizards can assist one another as well as define and modify an array of cognitively and physically demanding tasks on demand.} We thus developed a system to explore:

\sy{\begin{itemize}
    \item How Multiple Wizards collaborate while simulating a speech-based text input and editing task?
    \item How collaborative system functions might assist Multiple Wizards on developing collaborative strategies?
    \item How Wizards modify their collaborative strategies according to the end-user's responses?
\end{itemize}
}

\subsection{\sy{System Design of Wizundry 1.0}}
\begin{figure}
  \centering
  \begin{minipage}[t]{0.6\textwidth}
    \centering
    \includegraphics[width=\textwidth]{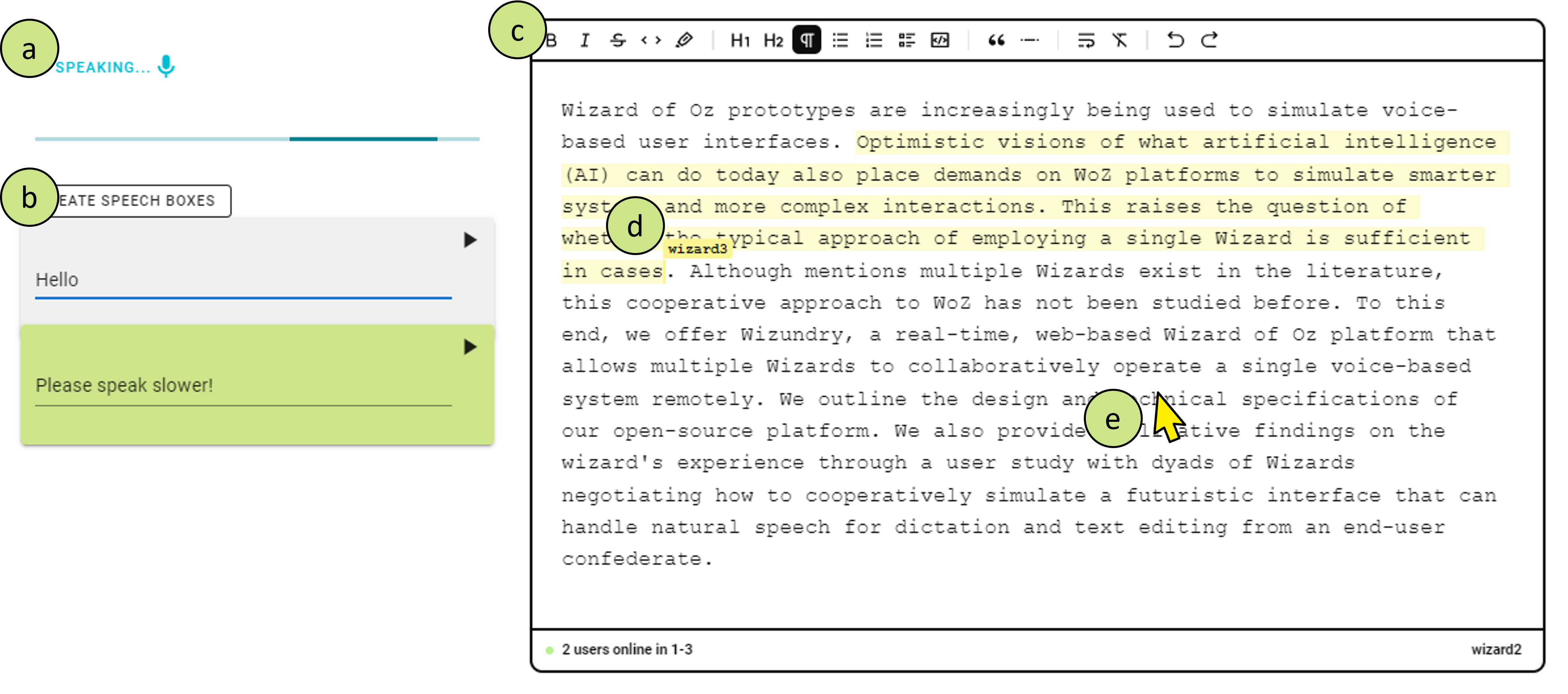}
    \caption{Wizards interface of Wizundry 1.0}
    \label{fig:Wizard-ui}
  \end{minipage}
  \begin{minipage}[t]{0.38\textwidth}
    \centering
    \includegraphics[width=\textwidth]{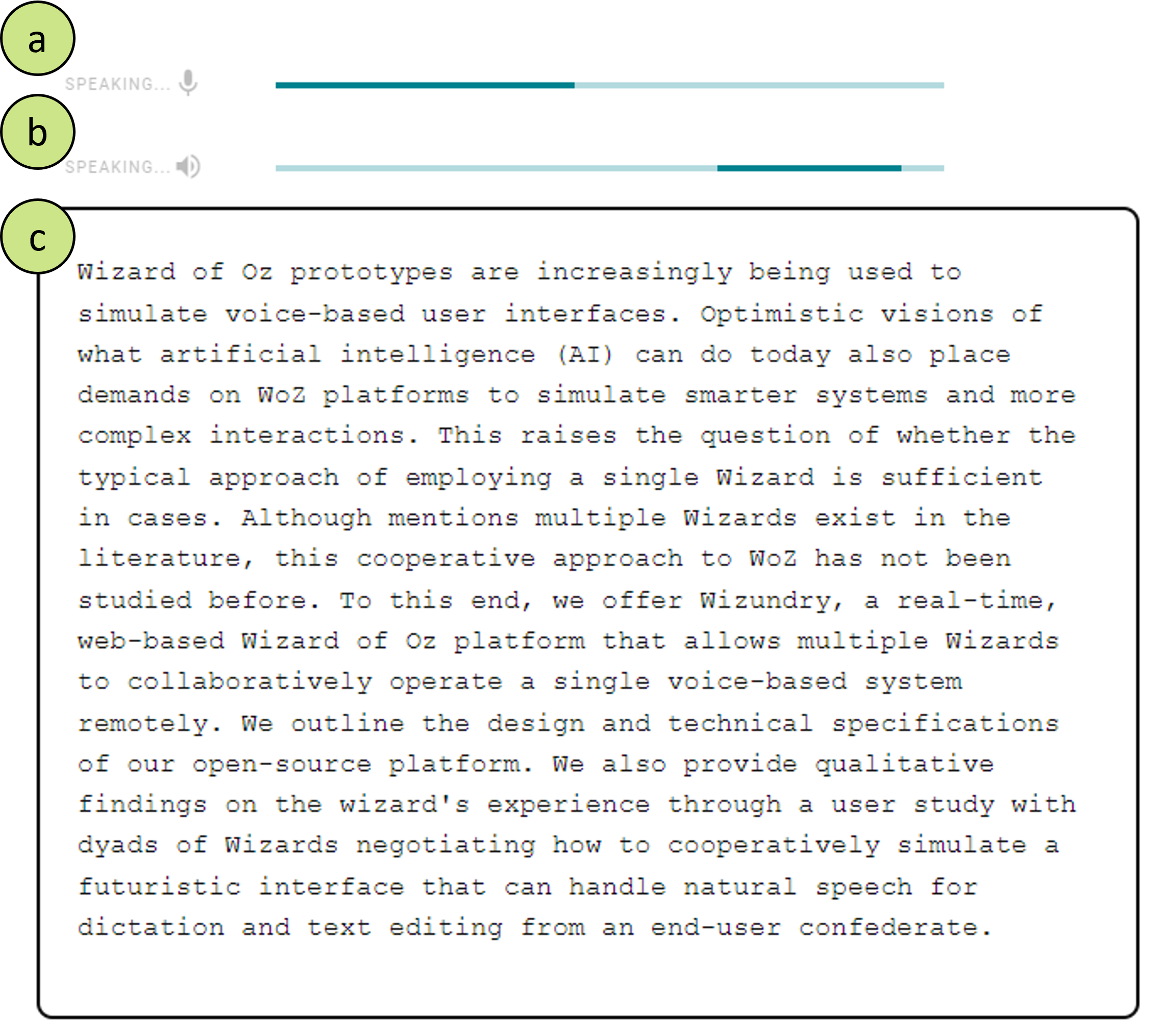}
    \caption{End-User interface of Wizundry 1.0}
    \label{fig:user-ui}
  \end{minipage}
\end{figure}

\sy{
We designed Wizundry 1.0 to allow researchers to (1) design and orchestrate speech-based WoZ studies with Multiple Wizards; (2) simulate speech-based intelligent features (such as keywords spotting and editing through natural speech); and (3) study how to provide better end-user experiences.}
\sy{The Wizundry consists of two interfaces: the Wizard interface (Fig \ref{fig:Wizard-ui}) and the end-user interface (Fig \ref{fig:user-ui}). The Wizard interface allows Wizards to (1) simulate a smart dictation interface by editing machine-transcribed speech text, (2) work collaboratively with others to accomplish tasks, and (3) test coordination workflows to improve the experience for end-users. The end-user interface is independent, and was used to test the simulated speech interface. This toolkit provides a modular and configurable interface that builds on two core functionalities: Speech-to-Text (STT) and Text-to-Speech (TTS), which can be used together or separately in each interface. }

\subsubsection{\sy{Wizard Interface.}}
\sy{The Wizard interface allows Wizards to edit dictated text, supports their awareness of each other's actions while collaborating together, and enhances the simulation of automated speech-based dictation and processing features.} We now describe the main features in detail.

\paragraph{\sy{Support collaborative editing of dictated text.}}
\sy{
A basic requirement entails transcription and text-processing mechanisms supported by collaborative writing functions. In Wizundry, 
this is achieved through the \textit{\emph{Collaborative Editor}} (Fig \ref{fig:Wizard-ui}.d), which all Wizards have access to. Inspired by Google Docs \footnote{https://www.google.com/docs/about/}, Wizundry allows end-user speech to be transcribed into text in the Collaborative Editor, where Multiple Wizards can edit the text content simultaneously.}

\paragraph{\sy{Support Wizards' awareness of each other in collaboration.}}
\sy{
Multiple Wizards must be able to edit text and operate dictation functions simultaneously. With parallel tasks and potential work conflicts (e.g., modifying the same transcribed text) that are likely to occur, Wizards need to keep track of the work status of their peers in real-time. Wizundry should therefore provide information to enhance Wizards' collaborative awareness \cite{erickson2000social,li2017two}.
To this end, Wizundry was designed to provide transparency of each Wizard's actions through
a \textit{\emph{Transparency View of each Wizard's actions}} feature and \textit{\emph{Line Break}} feature. The Transparency View includes (1) a Collaborative Cursor (Fig \ref{fig:Wizard-ui}.e), which indicates the cursor position of each Wizard, and (2) a Name Flag (Fig \ref{fig:Wizard-ui}.d), which labels each Wizard by name as a pop-up over their cursor. Thus, each Wizard can see the movements of their peers in real-time. The Line Break function was inspired by the collaborative writing feature in LaTeX documents \footnote{https://www.latex-project.org/}. This function can automatically separate each line, allowing end-users and Wizards to align with the location to edit quickly.} 

\paragraph{\sy{Control of speech-based intelligent interactive features.}}
\sy{
The Wizard interface provides two interactive features: (1) \textit{\emph{control of end-user speech}}, especially control of the end-user's microphone (Fig \ref{fig:Wizard-ui}.a), and (2) \textit{\emph{control of system audio}}, or control  of the speech boxes for playing preset responses and generating verbal feedback to end-users (Fig \ref{fig:Wizard-ui}.b). Microphone control allows Wizards to prevent unintentional STT transcription of speech from end-users (e.g., self-repair utterances, redundant dictations).
Microphone state is synced among all Wizards. The speech boxes are editable and can be played to generate system speech through TTS, allowing Wizards to send preset or custom feedback to end-users quickly. Specifically, the white boxes operate like instant messaging fields in chat apps and will clear after the content is sent. The green boxes hold preset messages that Wizards might frequently use and can be customized and repeatedly played during the study. Examples of the green box messages include ``Well noted!'', ``Can you speak slowly?'' Both speech boxes can be played by clicking the play button or hitting ``Enter'' on the keyboard when they are selected.}

\paragraph{\sy{Support flexible, varying workflows.}} 
\sy{
Drawing from the lightweight write-locking mechanism in Apparition \cite{lasecki2015apparition}, 
Wizundry allows Wizards to flexibly self-allocate tasks, try out collaborative workflows, and receive end-user feedback. Wizards can negotiate how to use the system to satisfy the end-user. The results of each Wizard's operation are synchronized to the end-user side. Notably, Wizundry is not prescriptive, allowing Wizards to dynamically and flexibly decide when and how to use the interface to collaborate or work alone, depending on the responses of the end-user.}

\subsubsection{\sy{End-User Interface.}}
\sy{This interface is what the end-user experiences. From an end-user perspective, it allows the end-user to interact with the system, which is simulated by one or more Wizards. From the Wizard's perspective, it is a way of checking the results of their performance and receiving commands from the end-user. From a research perspective, it allows for evaluations of the end-user experience as determined by the performance of the Wizards. We now describe the main features.}

\paragraph{\sy{Provide smart dictation service.}}
\sy{The \textit{\emph{Smart Dictator}} feature aims to simulate (1) transcription of the end-user's speech into text, (2) parsing of natural speech editing requests from the end-user, and (3) editing of dictated text. Drawing inspiration from the key transcription functions of dictation applications such as Otter.ai \footnote{https://otter.ai/individuals} and video auto-caption generation system iTour \footnote{https://www.itourtranslator.com/}, the end-user interface provides a text editor that shows the dictated-text input and editing results in real-time (refer to Fig \ref{fig:user-ui}.c).} 

\paragraph{\sy{Provide visual cues of system status.}}
\sy{The \textit{\emph{Speech/Audio Indicator}} provides real-time status feedback on the microphone and speaker for end-users. The end-user can view the state of the microphone (Fig \ref{fig:user-ui}.a) and speaker (Fig \ref{fig:user-ui}.b) to determine whether or not the system is listening or speaking.}

\subsubsection{\sy{Implementation.}}
Wizundry is implemented as a web-based system powered by NodeJS, which can be accessed on desktop and mobile devices. All system's status and content are shared synchronously to every users in real-time with Socket.io. It uses the Mozilla Web Speech API\footnote{https://developer.mozilla.org/en-US/docs/Web/API/Web\_Speech\_API} for web streaming users' audio input and Text-To-Speech for ``Speech Boxes''. Wizundry 1.0 uses speech-to-text technology to accomplish real-time transcription while transcribing streaming audio chunks using the Google Cloud Speech API \footnote{https://cloud.google.com/speech-to-text}. The collaborative editor was built on top of TipTap\footnote{https://www.tiptap.dev/} and Yjs\footnote{https://yjs.dev/}.

\subsection{Wizundry 1.0 User Study: Evaluating Single vs. Multiple Wizards}

We conducted a user study to test the initial design of Wizundry 1.0 and get a preliminary understanding of the Multi-Wizard experience. A pilot study with two groups (G1 and G2) and a main study with four groups (G3-6) were conducted. We were especially interested in their strategies and cognitive workload, one of the key human factors hinted at in previous work. To this end, we: 1) compared Wizard behaviour in solo- and multi-Wizard conditions; 2) observed the strategies that Wizards naturally employed while performing in pairs; and 3) assessed the usability, UX, and subjective cognitive workload.

\subsubsection{Study Design}
We used a within-subject design with one condition: Number of Wizards (Single-Wizard or Wizard-Dyad). In the Single-Wizard condition, one participant took on the role of a Wizard by themselves. In the Wizard-Dyad condition, two participants were randomly paired to play the role of Wizards as a duo. 

\subsubsection{Main Task}
The main task for the Wizards was to use the Wizundry system to simulate an intelligent dictation interface by responding to the End-User's dictated speech and requests for text editing in real-time. In the Wizard-Dyad condition, the Wizards needed to work together to decide upon and optimize a collaborative approach to their Wizardry. They had time to work out a plan after a training session and before the End-User arrived. They were also able to talk freely during the main task.

On the End-User side, the main task was performed by a confederate, a researcher pretending to be an End-User. The confederate used prepared scripts to act out the End-User dictation tasks, which required both composing and editing. These scripts used dialogue from established natural speech authoring actions, such as re-speaking and self-repair ~\cite{10.3115/981574.981581,liu2006enriching}, or issuing commands \cite{ghosh2020commanding}. In line with natural speech for dictation and text editing ~\cite{alto1989experimenting,ghosh2020commanding}, the scripts involved a combination of Editing-after-Composition (EAC) strategies (e.g., ``stars, don’t need `stars', delete `stars' '') and Editing-while-Composing (EWC) strategies, such as re-speaking for overwriting. 

The confederate was trained and asked not to deviate from the scripts, so as to maintain a high degree of consistency across sessions.

\subsubsection{Apparatus}
The experiment was conducted in a hybrid setting, a mix of online and offline modes, with the Wizards, the End-User confederate, and a remote experimenter in the same virtual meeting session via Zoom. 

The Wizards participated mainly offline, using separate computers (ThinkPad X390 laptops) but sitting together in the same room as the ``experimental'' laptop (a 2019 13-inch MacBook Pro), which held the Zoom session. The Wizards were asked to disable their camera so as to maintain the illusion of interacting with a real End-User. Video and audio were recorded in Zoom as well as using screen recording software and an external camera, in case of Internet trouble.

\subsubsection{Participants}
We recruited 12 participants (4 women and 8 men) from a local university to form 6 randomly assigned Wizard dyads. The participants were strangers to each other. None had professional dictation training or WoZ experience. All were non-native, yet fluent, English speakers(IELTS score of 6.5+). 

\subsubsection{Procedure}
The experimenter first introduced the study, collected informed consent, and explained the settings of the Wizundry system to all participant/s. 

The participants began with a training session so as to get familiar with the role of the Wizard, the Wizundry interface, and the task. Then they started the trials with a remote End-User confederate. Each trial required the participants to compose and edit one piece of text according to the acting End-User's script. Two scripts were randomly chosen from the script repository and counterbalanced to ensure that different scripts were provided for each trial within one experiment.
The first two groups were treated as pilot studies for a small iteration of system features and study design. In the pilot, participants (in G1 and G2) completed two [Single-Wizard] and four [Dyad-Wizard] trials (in G3-G6). In the formal study, participants completed one [Single-Wizard] trial and four [Dyad-Wizard] trials.

Each Wizard participant began in the [Single-Wizard] condition, then paired up with another Wizard participant to form a dyad. 

In all trials, they were allowed to take as long as needed to complete the task.
After finishing all trials, the experimenter conducted a semi-structured interview with each dyad and asked individual participants to fill in the NASA-TLX questionnaire. 
The experiment took 2 hours for each dyad, with a break in the middle.

\subsubsection{Instruments, Measures, Data Collection, and Analysis}

The following data were collected: 1) screen recordings of participants and experimenters; 2) audio of participants' conversations; 3) audio recordings of participants' interviews; 4) observations of experimenters during the process; 5) NASA-TLX \cite{hart1988development} self-reports in a post-questionnaire; and 6) system log data, which logged the Wizard's usage of the Wizundry application components. 

Quantitative analysis was used for the operational log data recorded by the system. In line with Hart ~\cite{hart2006nasa}, we did not weight the NASA-TLX scores. Qualitative analysis was used to explore patterns in behaviours and attitudes in the observational and interview data, and thematic analysis was used to find meaningful patterns in the data.

\subsection{Findings from User Study with Wizundry 1.0}
Wizards were able to develop, employ, and modify their own approaches and strategies on the fly within the dyad-Wizard context of Wizundry 1.0. Dyads were able to freely define and agree on a workable approach to text- and audio-based Wizardry. Much of the current literature (e.g., \cite{10.1145/354401.354406,10.1145/3202185.3202207,8673205,10.1145/2909824.3020256,7745134}) has focused on how a single Wizard conducts WoZ experiments or on the various ways designers can reduce the Wizard's workload. In our study, we emphasized observing how participants create and adapt within dynamic and complex, yet supportive and collaborative Wizarding contexts. We first present initial findings from the study, and then describe the revisions we subsequently made to Wizundry and the study procedures. 

\subsubsection{Feedback from the Pilot Study}

A key challenge for Wizards was how to deal with \textit{search content that needs to be modified}. 
Wizards spent a lot of time searching for content that needed to be edited during the [Wizard-Dyad] condition. The rhythm of cooperation and the speed of task completion was thus disrupted. 

As G1P2 explained, ``I have not even found [where the End-User is at] yet, and he is already moving on to the next composition.'' The cognitive burden of having to decide and act at the same time was high. G2P2 said that ``there was no time to think about using the `Speech Boxes' function''.

Even so, we noticed that the other Wizard sometimes helped out by finding the editing location and indicating it by selecting the text in editor with name flag. We thus added the ``Collaborative Cursor'' function to Wizundry after the pilot study. 

\subsubsection{Dyad-Wizard Cooperative Strategies}
\begin{figure}[htbp]
    \centering
        \begin{minipage}[t]{0.48\textwidth}
        \centering
        \includegraphics[width=6cm]{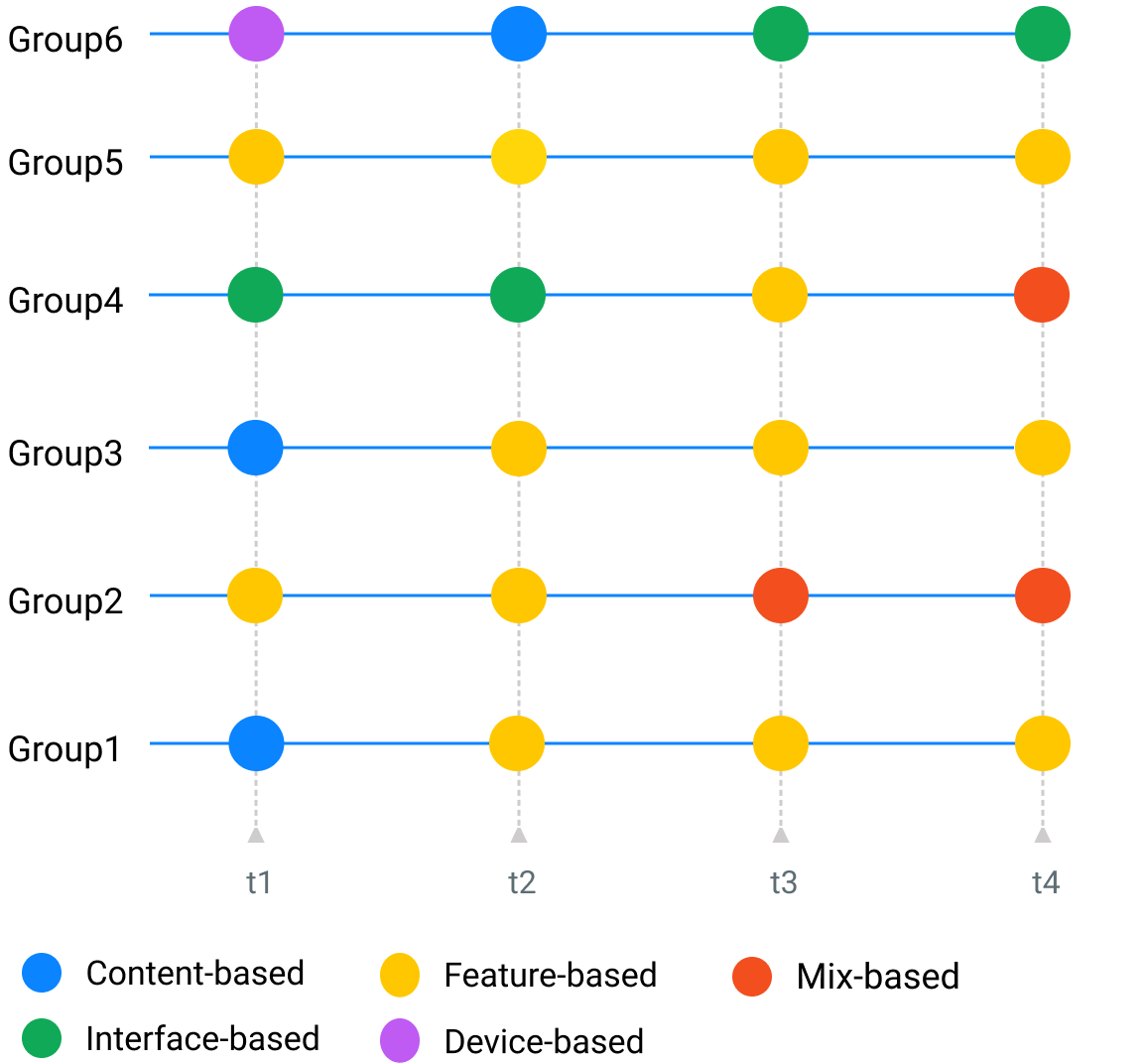}
        \caption{\sy{Group strategies for each of the [Dyad-Wizard] condition trials; The horizontal axis indicates the trial order and the vertical axis indicates the different groups; The various coloured dots represent the strategies used by each group in the four trials;}}
        \label{fig.groupstrategy}
    \end{minipage}
    \hfill 
        \begin{minipage}[t]{0.48\textwidth}
        \centering
        \includegraphics[width=6cm]{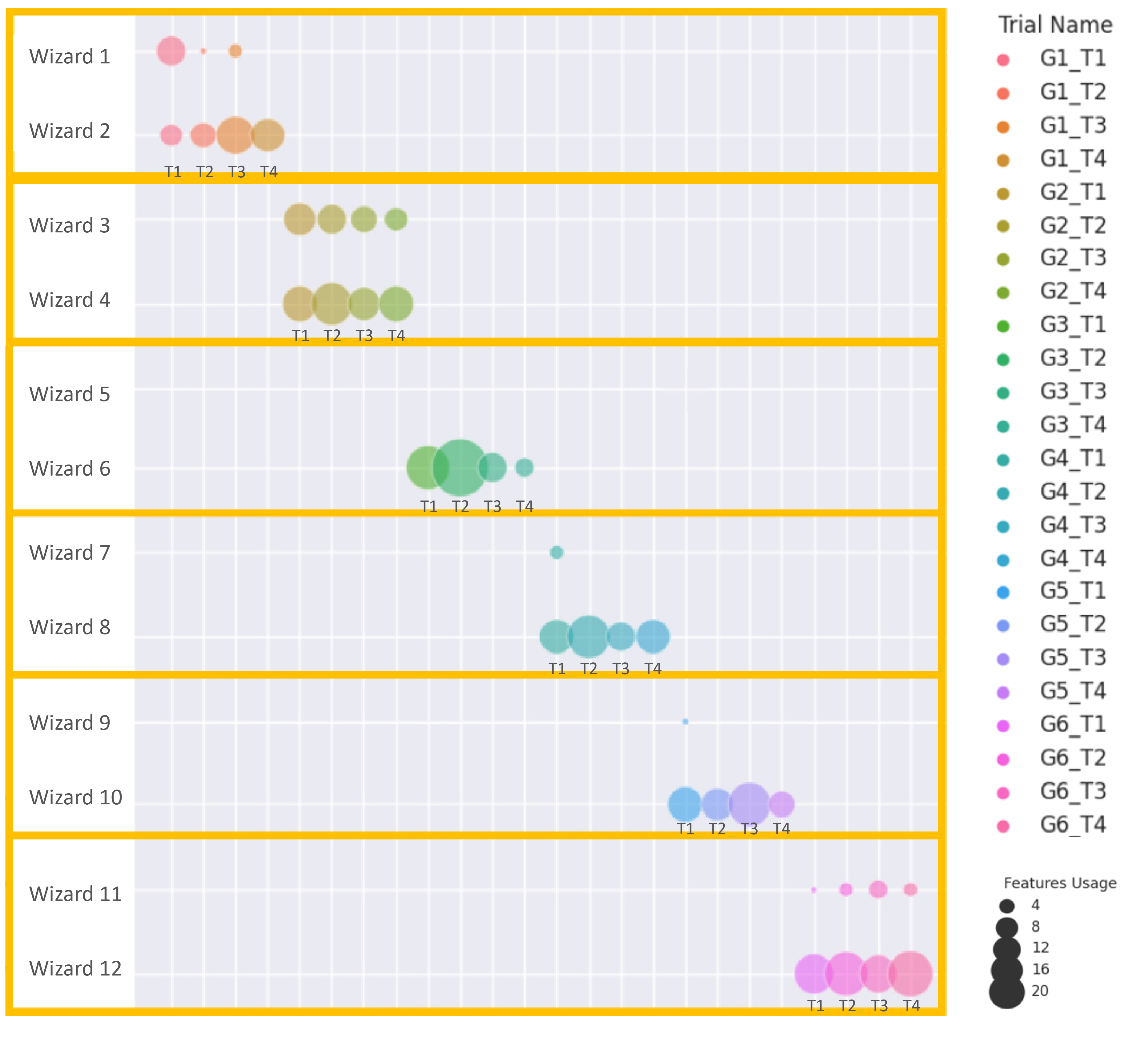}
        \caption{\sy{Visualisation of Wizard operation in each trial;The horizontal axis indicates the trial order and the vertical axis indicates the paired-Wizard for each group; The size of the dot represents the number of times the feature is used by Wizards;}}
        \label{fig.featureusage}
    \end{minipage}
\end{figure}

We observed \sy{five} types of cooperative strategies that can be categorized by the division of work between the Wizards. Five dyads altered the way that they worked collaboratively over the course of the study, while one dyad maintained the same approach from beginning to end. The strategies used in each dyad across trials, as shown in Fig.~\ref{fig.groupstrategy}, are described below.

\textit{Strategy: Content-based}.
G1, G3 and G6 adopted this strategy, with one Wizard responsible for composing and the other responsible for editing it when needed.
Wizards experienced significant issues with ``Microphone Control'' when using this strategy because there was no prior discussion about who would manage it. This resulted in all the speech of the End-User being dictated into the editor, thus leading to confusion in the distribution of work. As G1P1 explained, ``when the user speaks, the system mixes all the content together, with repeats and overlays, making it very difficult to tell what to edit out and what to [retain].'' 

\textit{Strategy: Interface-based}.
G4 and G6 divided their roles based on the interface areas in which the Wizards worked. Participants were responsible for either the left or right side of the interface. The area on the left contained the ``Microphone Control'' and ``Speech Boxes.'' The area on the right contained the Editor. G4 adopted this strategy after the first experiment, whereas G6 took it up for their last two trials, consecutively (see G4T1, G6T3/T4). The G4 Wizards could not handle both functions on the left side, so they switched their strategy in the remaining trials. G6 chose to follow this strategy until the end of the experiment.

\textit{\sy{Strategy: Feature-based}.}
\sy{This strategy involved one participant operating the ``Microphone Control'' or ``Speech Boxes'' functions while the other assumed the rest of the work. Whereas the Interface-based Strategy resulted from spatial and layout awareness, the Feature-based Strategy was developed due to Wizards' consideration of the system's functionality. The exclusive ``Microphone Control'' approach was used continuously in the last three trials by both G3 and G4 until the end of the experiment. We found that the Wizard controlling the microphone performed extra work spontaneously to help others. G3P6, for instance, provided additional reminders to the other Wizard based on the editing instructions issued by the End-User. This Wizard would proactively assist their teammate in finding the location of the text that the End-User needed to edit. They also actively helped remove extra dictation from the edit box. ``Speech Box-based'' approach, wherein one Wizard took full charge of the speech-box function, was our experiment's most frequently used strategy. We found that the timing in terms of switching the microphone on and off was most precise for the dyads using this strategy. However, making edits was relatively slow. One notable example is G5, which did not change its strategy after it was established at the beginning of the experiment. However, they communicated most frequently during the experiment compared to other dyads. In particular, the participant who was primarily engaged in ``Microphone'' and ``Speech Boxes'' control also helped the other Wizard locate edits.}

\textit{Strategy: Mixed}.
For G2, allocation of work was ambiguous and involved redundant working practices and ineffective results. We observed random jumps in the editing location of both Wizards which resulted in editing conflicts.  
According to G2P4, who was originally working on the microphone, ``I saw that [the other Wizard] was busy correcting something else and probably didn't notice, so I thought I would help him do it. But I thought it wasn't my job so I left it there.'' But his teammate commented, ``I saw [the other Wizard's] edit cursor was in that spot and thought he would help me directly, but he just stayed there. I was afraid we'd both change [something] at the same time.'' This shows that having unspoken expectations of each other may also create confusion.


\textit{Strategy: Device-based}.
One dyad divided the labour by selecting which hardware devices each controlled (see G6T1 in Fig~\ref{fig.groupstrategy}). One Wizard operated the mouse while the other used the keyboard. These Wizards felt that focusing on one device would reduce the burden of operation and improve efficiency. As G6P11 explained: ``I could click several buttons with the mouse, so I didn't have to press the keyboard with one hand and controlled the mouse at the same time.'' \sy{It is evident that the yellow dot representing the Feature-based strategy is the most frequently used form of Dyad-Wizard cooperation. Meanwhile Device-based strategy, represented in purple, is the least used strategy. Group 5 was the only group to maintain the Feature-based strategy while other groups changed their approaches to coordination.}

\textit{\sy{Spontaneous change of plan.}}
\sy{Each Wizard's usage of Wizundry's functions are visualized in Fig.~\ref{fig.featureusage}. 
We calculated the log data and number of operations of the Wizard in the system. In terms of "Features Usage", the size of the circles indicate the frequency of using ``Microphone Control'' and ``Speech Boxes'' in the study. Text editing actions performed by Wizards were not counted. Hence, the figure shows how many times each Wizard used features in each trial. The more a Wizard uses a feature, the larger its dotted area. It is also affected by the division of labour strategy.} 
The disparity in the actual and projected divisions of labor appeared to significantly affect performance. Before the session, some dyads agreed to distribute labor in accordance with the Mic-Speech cooperative strategy. However, examination of the logs revealed that the anticipated division of labor was not followed-through (see G1 and G2 in Fig.~\ref{fig.featureusage}). In contrast, the dyads who maintained consistency in the planned, and actual  execution of tasks performed well and considered the Dyad-Wizard environment to be successful and effortles (see G3 and G5 in Fig.~\ref{fig.featureusage}).
\sy{Group 3 initially adopted the Content-based approach and later followed the Feature-based strategy, which assigned Wizard 5 to focus on text modifications and wizard 6 to other tasks. Thus, the diagram shows that Wizard 5 has no record of operating on other features. Meanwhile, Group 4 and 5 only adapt the cooperative strategy and work overlap in the first trial, and later there is a clear division of labour. However, other groups, such as G2, have records of two wizards simultaneously operating functions even with a well-defined strategy of the division of labour.
Although the aim for not splitting up the job as intended was to increase efficiency, it resulted in an ambiguous division of labor, which was detrimental to the collaboration.}

\subsubsection{Mixed feelings about Dyad-Wizard}
Feelings about the Wizarding modes were mixed. Some felt that dyads were better (n=5) as their Dyads were effective at solving the difficulties encountered in solo mode. They elaborated that adding another Wizard could facilitate better control of the system and accomplish the Wizard's goals. As G3P5 said, ``two Wizards [led to] a better outcome and efficiency, which could let [us] focus on [our] own work.'' Others felt that adding a teammate reduced their workload by offloading tasks, especially helping them to control the microphone and by being able to use the ``Speech Boxes'' function. As G4P8 said, ``I wouldn't have used the speech boxes function before because I didn't have time to look at it.''

On the other hand, others felt that a single Wizard was sufficient for the task and that a additional Wizard was of limited benefit (n=4), whereas the rest felt that aligning with an extra person was in fact disruptive to their task at hand(n=3). Some participants expressed that the other Wizard simply replicated what they could already do alone, and that this affected their own performance to the extent that they felt better off in the [Single-Wizard] condition. As G2P3 explained, ``I could do it myself. It's a waste of time when I had one more person to think about.''. Even in dyads with better outcomes, some participants felt that one Wizard was enough. For example, G1P2 said, ``What I could do it by myself, one more person would only hinder me.''

\subsubsection{Dominant Personalities and Performance}
In G3, G4, and G5, the presence of a dominant Wizard reduced the occurrence of conflicts and led to both Wizards finishing the job efficiently. The dominant participant allocated the work while the other Wizard acted as the co-pilot. 

The confederate End-User noted that these dyads completed their tasks faster than others. This also improved the End-User experience, especially for prompt feedback about their requests. Most dominant Wizards controlled the ``Microphone'' and spoke more. Without prior consultation, they were often able to quickly provide their teammates with a place to edit the text. As G4P8, a dominant Wizard, said, ``I noticed that [the other Wizard] didn't seem to hear what the user had just said, so I reminded him.'' We also observed ``code-switching'', where Wizards switched from English to their native language to quickly pass on time-sensitive or complex information.

\subsubsection {Attitudes towards System Features}
Most Wizards rarely used the ``Speech Boxes'' function, even though they were responsible for controlling it. They felt that they were in control based on the End-User's instructions and so did not need to confirm with them. When in solo mode, Wizards reported that they had no capacity to mouse click on the play button. The Collaborative Cursor required extra effort and time to look for other Wizards' position and to ascertain their current actions. Moreover, most Wizards were not aware of the line break function developed to support Multiple Wizards in writing the content together, so they overlooked it even though they had been trained to use it. Nevertheless, several participants found this function extremely useful, especially for navigating editing instructions from the End-User (G5).

\subsubsection{Cognitive Workload}
The raw, summed, and averaged NASA-TLX scores across participants for the dyad context are presented in Appendix Table~\ref{tab:nasatlx}. Participants in the [Dyad-Wizard] condition had an overall average summed score of 31.9 (SD=7.1, MED=31, IQR=19) and mean score of 5.3 (SD=1.2, MED=5.2, IQR=3.2). With respect to the dyadic context, we noticed that some participants went out of their way to help their partner, but this was often disruptive rather than helpful.

\subsection{Summary}
Wizards were able to tackle a variety of challenges as they experimented with the various novel strategies to collaborate in Wizundry 1.0. 
Most dyads focused on how to handle the demands of the WoZ dictation task, but neglected teamwork as a result. In addition, communication and personality factors influenced the dynamic within the dyad, which then affected performance. When unexpected events happened outside the established cooperative approach, some participants initiated communication with the other Wizard, positioning themselves as the leader or dominant personality. Often this led to performance improvements, with the dominant personality proactively changing their behaviour as needed. In short, Wizundry 1.0 was easy to use in most cases, and the addition of a partner could be useful.


\section{Prototyping and Evaluating Wizundry 2.0}
With Wizundry 1.0, we created a basic platform for simulating intelligent dictation interface features and evaluated its effectiveness with solo Wizards and dyads. Now we describe how we iterated the design from 1.0 to 2.0. Then, we report on a second user study involving three Wizards triads who simulated more advanced intelligence features. We start with our iterative design process for Wizundry 2.0.

\subsection{\sy{Iteration and System Design of Wizundry 2.0}}
\begin{figure}[b]
  \centering
  \includegraphics[width=1.0\linewidth]{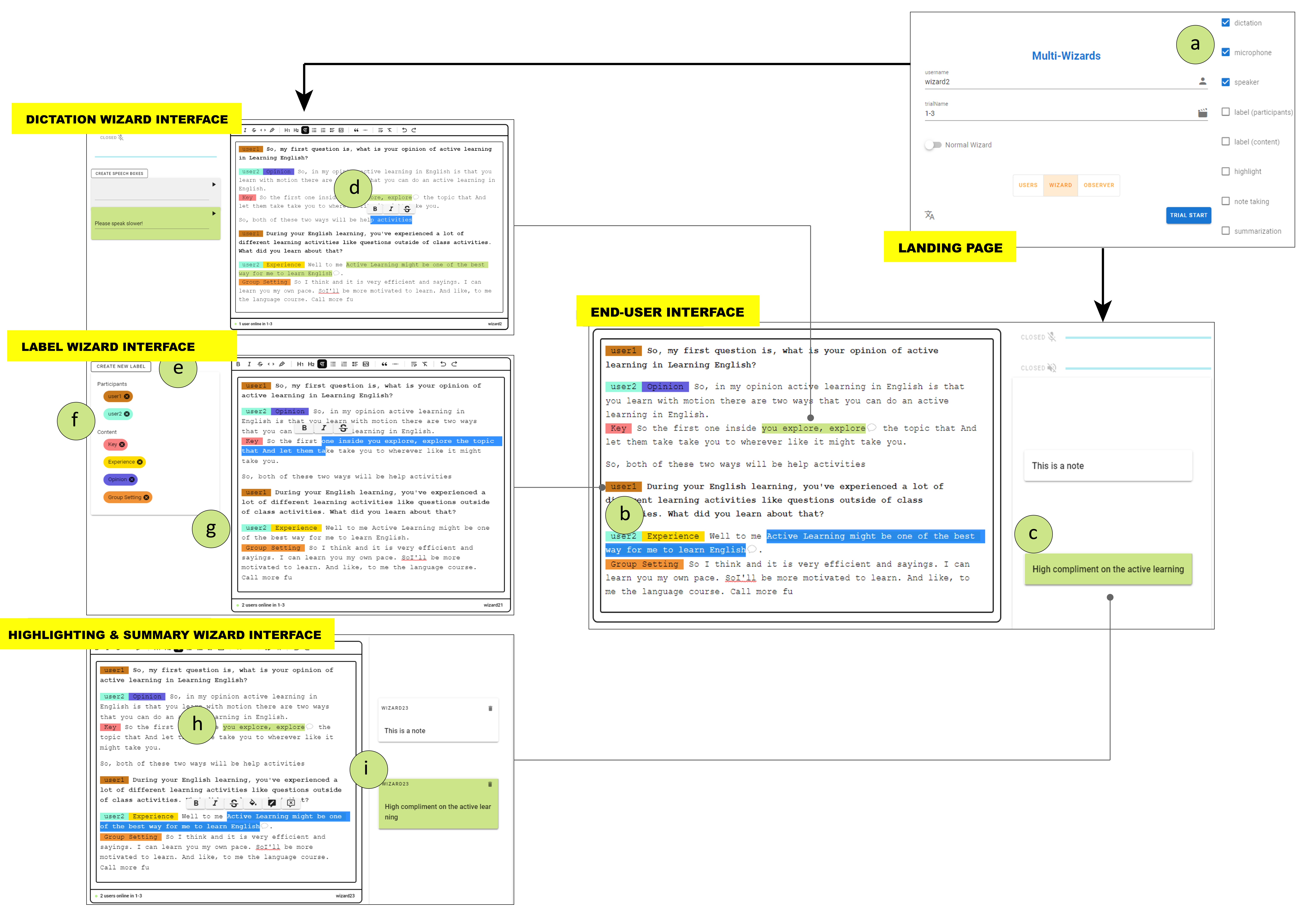}
  \caption{Wizundry 2.0 features and role-based interfaces; In Landing Page: a) features selection; In End-User Interface: b) labeled outcome; c) note taking result; In Dictation Wizard (DW) Interface: d) edit STT result; In Labelling Wizard (LW) Interface: e) new label creation; f) created labels; g) labelling on text; In Highlight with Summary Wizard (HSW) Interface: h) noting effect on text; i) bubble menu with highlight, add and delete note; c) related note to the selected text.}
  \label{fig:w2}
\end{figure}

\sy{
\textit{Wizundry 2.0} includes three interfaces: (1) the Landing Interface, (2) the Wizard Interface, and (3) the End-user Interface (see Fig \ref{fig:w2}). Here, we describe how we translated our findings into design goals and then implemented them in Wizundry 2.0.}

\subsubsection{\sy{Design Goals and Additional Features in Wizundry 2.0}}

\paragraph{\sy{Facilitate Wizards' capacity to handle real-time demands.}}
\sy{Wizundry 1.0 provided a set of standard editing tool (e.g. bold, italic, and headers), but Wizards could not use it due to the rapid pace of interaction. The operation distance from the toolbar on the editor to the latest displayed text was too far. Inspired by the pop-up toolbar designed for synchronized proofreading task with crowd-worker in Soylent \cite{bernstein2010soylent}, Wizundry 2.0 remedied this gap by adding a new feature: the \textit{\emph{Bubble Menu option}} (Fig \ref{fig:w2}.d). When Wizards select text, a menu would appear above the mouse, allowing for faster selection of frequently used functions.}

\paragraph{\sy{Support the simulation of advanced dictation.}}
\sy{
Wizundry 2.0 needed to support the prototyping of more advanced dictation features based on intelligent text processing outputs, such as real-time close captioning from live-streaming. 
Recent advancements in NLP have made attempts to support more advanced semantic analysis and information extraction \cite{paulheim2017knowledge}, which are increasingly utilized in HCI~\cite{cheng-etal-2022-mapping}. Wizundry 2.0 takes such features as examples of intelligent features to be used in a future dictation system, by providing the following functionality: (1) \textit{\emph{Text Summary}}, (2) \textit{\emph{Text Highlighting}}, and (3) \textit{\emph{Sentence-based Labelling}}. }

As in multilevel information sharing between workers in Cobi and StructFeed \cite{huang2017supporting, kim2013cobi}, the Text Highlighting feature was expanded with various background colours so that Wizards could easily discern among keywords. Following the research on providing feedback summaries during collaborative writing tasks \cite{huang2017supporting}, we added the Text Summary feature (Fig \ref{fig:w2}.h), which can select text content and add relevant information into the working area (Fig \ref{fig:w2}.i). Wizards can also select an entire sentence or multiple lines for labelling (Fig \ref{fig:w2}.g) and self-defined labels are automatically added to the header position. This enables Wizards to add or remove labels (Fig \ref{fig:w2}.e) and make label categories (Fig \ref{fig:w2}.f) before the study begins. These modifications are synchronized to the end-user interface, as shown in Fig \ref{fig:w2}.b and Fig \ref{fig:w2}.c.

\paragraph{\sy{Provide advanced dictation context services.}}
\sy{
Speed reading and more advanced text processing are desired by end-users. As such, we added a \textit{\emph{Tailored dictator}} feature to provide dictation context services based on real-time end-user needs. This is especially pertinent when the end-user is performing an interviewer task: automatic keyword spotting, categorizing different sentences with labels, and providing content summaries. 
It also can be used to obtain design insights and to test various types of information processing in WoZ tasks.}

\paragraph{\sy{Enable feature combinations and independent work.}}
\sy{We discovered that Wizards adapted their strategies quickly and innovatively to meet end-user demands in the Wizundry 1.0 study. Thus, we ensured that Wizundry 2.0 enables a clear division of labour so that Wizards can focus on their own interface and tasks. Inspired by work on how to avoid collaborative conflicts in crowdsourcing platforms \cite{bernstein2010soylent}, we added a new interface: the Landing Interface (top of \ref{fig:w2}). The interface is modular, enabling the combination of various features (e.g., Collaboration Editor and Text Highlighting) flexibly into each Wizard's respective interface. There are two main features: (1) \textit{\emph{Feature Selection List}} and (2) \textit{\emph{System User assignment}} in the landing interface. Researchers can then assign each Wizard their own tasks or distribute independent functions in the Wizard's interface before the study, while also allowing Wizards to rapidly adapt their workflow during the study.}

\paragraph{\sy{Technical advances.}}
Lastly, we leveraged the capabilities of the Vue.js front-end framework to enable future developers to create new features in a component-based approach.

\subsection{Evaluating Wizundry 2.0: Understanding the Collaboration between End-Users and Wizards}

We conducted a case study to evaluate Wizundry 2.0 with three Wizards collaborating together to simulate more advanced intelligent system features in addition to dictation. We targeted an imagined future dictation system that supports time-critical conversational tasks, such as an interviewer managing an interview process or debaters engaging in debates using agendas. Traditional dictation software cannot support such time-critical tasks well because it takes time to read lengthy transcriptions that have no structure, summaries, or highlights.  

For our case study, we chose an \textit{interview speech notes processing task} because it is complex enough to require multiple Wizards when simulated, as well as covers most transcription and content analysis requirements for research purposes, such as initial qualitative processing of transcribed text. We aimed to test whether and how Wizundry 2.0 can help researchers prototype and test future interfaces. We evaluated the effectiveness of our Multi-Wizard platform based on the following aspects: Can it help researchers simulate a near-future system that has advanced intelligent features, one that is hard to simulate by one Wizard alone? Can it support fast iterations of the simulated interface by adapting to the cooperative strategies of multiple Wizards? Can it help researchers learn how to improve the end-user experience? 
We describe the details of the study next.

\begin{figure}
  \centering
  \includegraphics[width=0.65\linewidth]{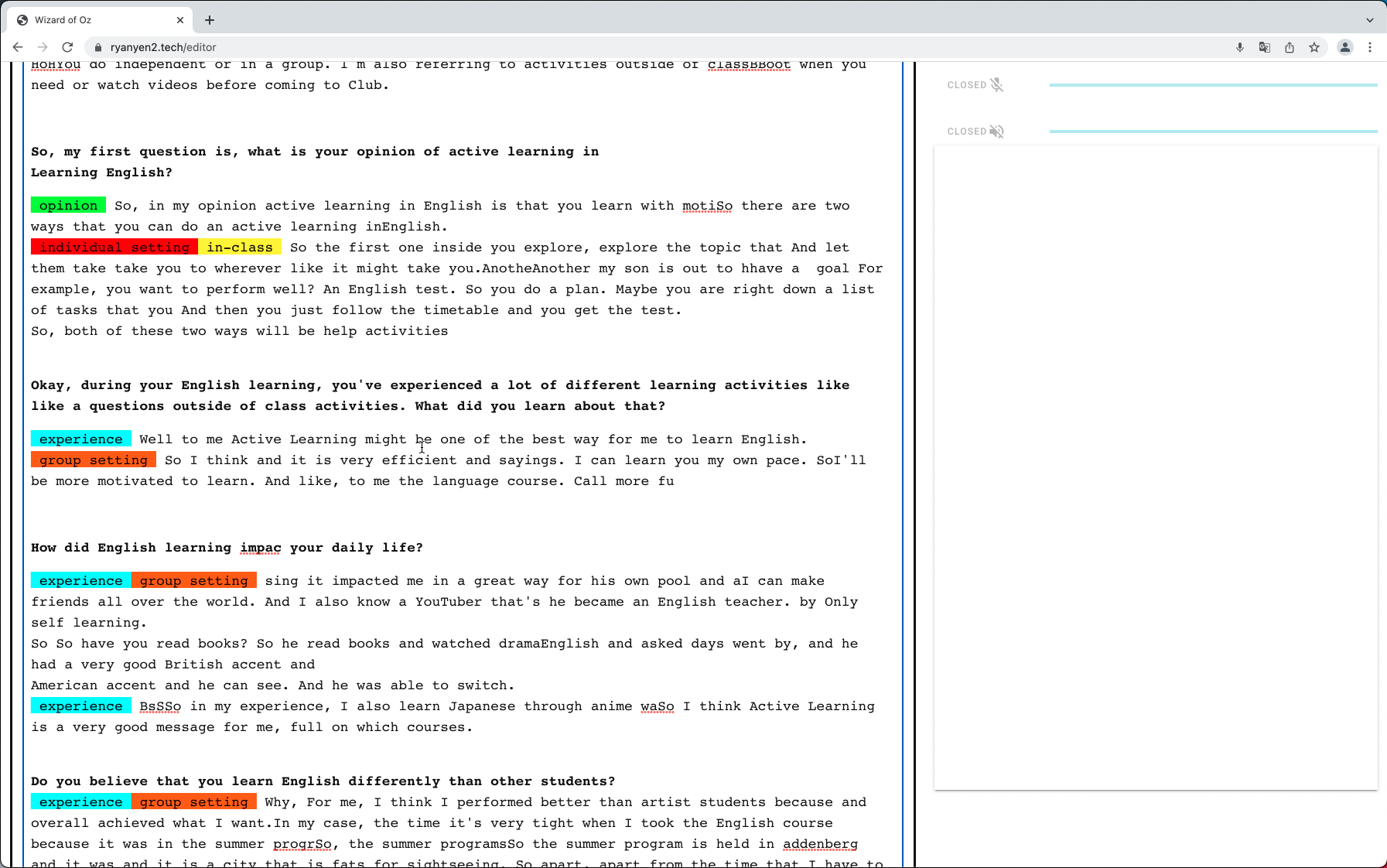}
  \caption{End-user interface while performing an interviewer task.}
  \label{fig.scriptexample}
\end{figure}

\subsubsection{Main Task and Roles}
The task simulates a situation wherein a semi-structured interview is conducted so as to generate insights about the interviewer and Multi-Wizard collaboration experiences when using AI-based dictation interfaces. For this, two trained confederates from the research team acted as End-Users, taking on the Interviewer and Interviewee roles in the task scenario for all the Wizard groups. 

Each group of Wizards consisted of three participants who worked on separate computers with a synchronized interface to complete tasks as a team. Participants needed to find their way to work together so as to simulate a unified intelligent dictation interface. Their main task for this was to provide interactive dictation services to End-Users per their requests. They were given the goal of trying to provide a good End-User experience with their simulation by responding to End-Users' requests timely and accurately while providing necessary system feedback. They were allowed to communicate freely whilst trying to accomplish the task as quickly as possible.

As shown in Fig \ref{fig.scriptexample}, in the context of the semi-structured interview, each Wizard was assigned to one of three different roles: Dictation, Labelling, or Highlighting and Summarizing: 

\begin{itemize}

\item[-] \underline{Dictation Wizard (DW)} is based on User study 1.0, with the same task design. A ``Dictation Wizard'' should act as a smart dictation interface, dictate and editing the interview content by following the interviewer's requests.
\item[-] \underline{Labelling Wizard (LW)} is assigned to structure the interview consensus. They were tasked to categorise interview transcripts into various dimensions according to interviewer's label list.
\item[-] \underline{Highlight and Summary Wizard (HSW)} needs to highlight key content and write summary notes. 

\end{itemize}

In order to maintain consistency, two researchers were trained as the Interviewer and Interviewee who would perform the activities according to prepared scripts (Appendix. tbd) and interview topics, as shown in Fig.\ref{fig.scriptexample}. 

\subsubsection{Location and Apparatus}
The studies were conducted in two separate soundproof rooms, with the Wizards, the End-Users, and the coordinating researcher connected through Zoom. This online meeting format allowed researchers to organize and control the voice communication between the two sides. On the Wizard's side, the Observer muted the microphone while listening to the audio of the end-user's speech.
Video and audio were recorded using an integrated screen recording feature
on all computers, as well as audio and video recorders as back-ups.

\subsubsection{Participants}
We recruited 12 participants (4 women and 8 men) from local universities to form four groups (Table.\ref{tab:Wizardid}). Tasks were completed in English. All participants studied in academic programs taught in English and scored a minimum of 6.5 on the IELTS. To make sure that the participants understood the goal of simulating a good End-User experience, we recruited PhD students who had research experience in diverse domains within HCI. The participants had no prior experience with WoZ studies.

\begin{table}[]
    \small
    \centering
    \begin{tabular}{cccc}
    \hline
    \textbf{Participant} &
      \textbf{Group} &
      \textbf{Gender} &
      \textbf{\begin{tabular}[c]{@{}c@{}}Wizard Role \end{tabular}} \\
      \hline
    P1  & \multirow{3}{*}{G1}  & F & Highlight and Summary  \\
    P2  &                      & M & Labelling              \\ 
    P3  &                      & M & Dictation              \\ \hline
    P4  & \multirow{3}{*}{G2}  & M & Labelling              \\
    P5  &                      & M & Dictation              \\ 
    P6  &                      & F & Highlight and Summary  \\ \hline
    P7  & \multirow{3}{*}{G3}  & M & Dictation              \\ 
    P8  &                      & F & Labelling              \\
    P9  &                      & M & Highlight and Summary  \\ \hline
    P10 & \multirow{3}{*}{G4}  & M & Highlight and Summary  \\ 
    P11 &                      & F & Labelling              \\
    P12 &                      & M & Dictation              \\ \hline
    \end{tabular} 
    \caption{Summary of participant demographic information and Wizard roles. No one changed roles between trials.}
    \vspace{-20px}
    \label{tab:Wizardid} 
    \setlength{\tabcolsep}{1mm}
\end{table}

\subsubsection{Procedure}
Before starting the session, the researcher introduced the study, collected informed consent, and demoed the whole system. After the training, each participant was familiarized with the features of the Wizard operation interface.  
Each Wizard was then assigned one of the three roles we described previously. Each participant was then provided with the End-User requirements for the interview task, such as label list, highlight categories, and summary requirements (see the Appendices for details). After a team discussion, the Wizards decided on their individual tasks and moved to the role-based interface (see Fig.\ref{fig:w2}).
They were asked to work collectively as a unified intelligent dictation system for the interviewer to use. The basic questions and structure of the semi-structured interviews conducted by the End-User remained consistent across trials and used similar topics (e.g., active learning or language learning). The order of topics was counterbalanced across trials. Next, the main task began, divided into three steps as below:

\begin{itemize}

\item[-] \textit{Step 1. The first trial:} Wizards begin the first trial by simulating the system \sy{for the first time}. 
\item[-] \textit{Step 2. Listen to End-User feedback and re-discuss strategy:} The End-User was asked to fill in the NASA-TLX and talk about their experience in terms of what worked and what did not. Wizards listened in to the End-User's feedback quietly via an audio communication channel. Afterwards, Wizards were asked to renegotiate their collaboration strategy to improve the End-User experience, for instance, by optimizing their division of labor and workflow to simulate features more in-line with the End-User's expectations. This was to test whether Wizards could quickly adapt to End-User requests and feedback so as to improve the End-User experience they could provide using our platform.

\item[-] \textit{Step 3. The second trial:} After agreeing on a modified strategy, the Wizards worked together again to simulate the system for a similar task. 

\end{itemize}

We  specified a high level of difficulty for the Wizard task by asking the End-Users to provide a list of basic services that they wished the simulated system would provide. We asked End-Users to do this in advance of the study based on our pilot tests, wherein we found that Wizards needed specific requirements to be able to provide effective help to the End-User. The End-Users' requested services were twofold: auto-tagging of a list of labels for each interview task, and auto-styling of text based on their requests to bold or highlight certain text in the transcribed text, such as the interview questions.

In all, each participant completed two trials as the Wizard. During the trials, there were no restrictions placed on communication between the Wizards, but they were asked to do their best to provide a good End-User experience even while communicating with each other. After completing all trials, Wizards and End-Users were required to complete the NASA-TLX as a measure of the experience of the system and cognitive workload, respectively. The researcher also conducted semi-structured interviews with everyone. These interviews explored the Wizard's collaboration, the Wizard's use of the system, and the End-Users' feedback on the simulated system.

\subsubsection{Data Collection and Analysis}
In total, we collected 3 Wizards x 2 trials x 4 groups = 24 trials worth of data on the Wizard side, and 1 interview x 2 trials x 4 groups = 8 trials worth of data on the End-User side. The data used for qualitative analysis consisted of transcribed audio recordings of interviews conducted with the Wizards and End-Users.
\sy{We used inductive thematic analysis to analyze this data so as to gain insights on each Wizard's perspective of the task, system usage, and team collaboration \cite{fereday2006demonstrating}. One researcher first conducted open coding of the transcripts and then grouped these into themes and sub-themes based on the interview questions.} 
The NASA-TLX data used for the quantitative analysis was captured in post-study questionnaires completed by each Wizard and the End-Users.

\subsection{Findings}
We first present the findings from the Wizard collaboration results, including four cooperation strategies and perceptions of the effectiveness of Multi-Wizard collaboration as a triad. We then cover reactions from the Wizards and End-Users who acted as the Interviewer and Interviewee. Finally, we review the interaction patterns that we observed in each group.

\subsubsection{Cooperation between Wizards}
All groups of Wizards tried to perform the three system features simultaneously for the first trial (see Initial collaboration strategy in Table.\ref{tab_multiWizardstrategy}). Every group adapted their strategy into their own after receiving End-User feedback, for enhancing output quality and work efficiency. 

We can summarize the collaboration strategies carried out by teams of Wizards in three categories: Sequential, Interruption, and Hybrid strategies. 
\sy{We provide an overview of the strategies used by each group in Table.\ref{tab_multiWizardstrategy} and their outcomes in Table. \ref{tab_strategyfeedback}. Each group of strategies corresponds to a summary of the End-Users' experiences resulting from the Wizards' cooperative work 
in the first trial and after the iteration phase, as well as each Wizard's perceptions and experiences when performing the task.}

\underline{The Pipeline Strategy} was mainly used in the first group (see ``Group One'' in Table.\ref{tab_multiWizardstrategy}). In this strategy, Wizards started with the dictation task and then the labelling, highlighting, and summarizing tasks, which were completed in sequence. Each Wizard's work built on the output of the other Wizards. 

\underline{The Coordinated Strategy} was mainly used in the second group (see ``Group Two'' in Table.\ref{tab_multiWizardstrategy}). In this strategy, the Dictation Wizard controlled the pace of the workflow by providing frequent verbal feedback to End-User via the speech boxes function. After experiencing some difficulties
in the first trial, Group Two then created multiple speech boxes to define feedback messages in advance, such as ``please wait a moment''. In the second trial, the Dictation Wizard used the speech boxes to interrupt the End-User whenever they detected that the other two Wizards could not catch up with the speed of the End-User. In a typical interruption scenario, the Dictation Wizard first confirmed that the Label Wizard and Highlight+Summary Wizard had completed their parts and then  asked the End-User to continue. 

\underline{The Spontaneous Strategy} emerged in the third and fourth groups (see ``Group Three and Four'' in Table.\ref{tab_multiWizardstrategy}). These groups were focused more on the segmentation of dictated text and had overlapping responsibilities as Dictation Wizard and Labelling Wizard. 
We observed that the Wizards in both groups spontaneously assisted each other's work. 
The text editing tasks were handled jointly between DW and HSW by whoever had the time and capacity in the moment. The Label Wizard and the Dictation Wizard typically edited and separated the text simultaneously. Then, the Wizard responsible for highlighting and summarizing proceeded based on the team consensus (see ``Group Three'' and ``Group Four'' in Table.\ref{tab_multiWizardstrategy}). Therefore, Group Three and Four adopted a hybrid strategy, mixing the Pipeline and Spontaneous Strategies.

\begin{table*}
    \scalebox{0.70}{
    \begin{tabular}{c|c|c|c|c|c|c}
        \hline
        \textbf{Strategy} & \textbf{Type} & \textbf{Sub-type} & \textbf{Group 1} & \textbf{Group 2} & \textbf{Group 3} & \textbf{Group 4} \\
        \hline
        \multirow{4}{*}{Initial Strategy}
        &Task division & Focusing on their assigned roles    & \ding{51} & \ding{51}  & \ding{51} & \ding{51} \\
        &Time division & Everyone worked in parallel    & \ding{51} & \ding{51}  & \ding{51} & \ding{51} \\
        &Relationship & Equal                           & \ding{51} & \ding{51}  & \ding{51} & \ding{51} \\
        &Dependency & Independent                       & \ding{51} & \ding{51}  & \ding{51} & \ding{51} \\
        \hline
        \multirow{15}{*}{Adapted Strategy} 
        & \multirow{3}{*}{Task division} & The Pipeline Strategy                    &\ding{51} & \ding{55} & \ding{51} & \ding{51} \\
        &                                & The Coordinated Strategy                  &\ding{55} & \ding{51} & \ding{55} & \ding{55} \\
        &                                & The Spontaneous Strategy                        &\ding{55} & \ding{55} & \ding{51} & \ding{51} \\
        \cline{2-7}
        & \multirow{3}{*}{Time division} & First DW then LW then HSW                  &\ding{51} & \ding{55} & \ding{55} & \ding{55} \\
        &                                & First DW then LW and HSW                   &\ding{55} & \ding{51} & \ding{55} & \ding{55} \\
        &                                & First DW and HSW then LW                   &\ding{55} & \ding{55} & \ding{51} & \ding{51} \\
        \cline{2-7}
        & \multirow{3}{*}{Relationship}  & Equal but separated execution                  &\ding{51} & \ding{55} & \ding{55} & \ding{55} \\
        &                                & Equal and helping each other      &\ding{55} & \ding{55} & \ding{51} & \ding{51} \\
        &                                & DW led coordination                    &\ding{55} & \ding{51} & \ding{55} & \ding{55} \\
        \cline{2-7}
        & \multirow{3}{*}{Dependency}    & Relied on the work output of the previous Wizard &\ding{51} & \ding{55} & \ding{55} & \ding{55} \\
        &                                & LW and HSW relied on DW's work                      &\ding{55} & \ding{51} & \ding{55} & \ding{55} \\
        &                                & LW relied on DW's and HSW's work                        &\ding{55} & \ding{55} & \ding{51} & \ding{51} \\
        \hline
    \end{tabular}
    }
    \caption{\sy{The Multi-Wizard cooperation strategies of the first trial (initial strategy) and the second trial (adapted strategy) after receiving End-User feedback.}} 
    \label{tab_multiWizardstrategy}
\end{table*}

\begin{table*}
    \scalebox{0.55}{
    \begin{tabular}{c|c|c|c}
        \hline
        \textbf{Group} & \textbf{\makecell[c]{End-user experience\\ with Initial Strategy}} & \textbf{\makecell[c]{End-user experience\\ with Adapted Strategy}} & \textbf{\makecell[c]{Wizard Perception\\ with Adapted Strategy}} \\
        \hline
        \textbf{{Group 1}}
        &\multirow{4}{*}{\makecell[l]{1.Unsatisfactory User Experience;\\2.Lack of feedback from the svstem;\\3.Time lag between input and result;\\4.Disagreement with returned result;\\5.User wants alternatives to speech-Input;}}
        &\makecell[l]{1.The system was not providing\\ the required information;\\2.No response, too slow;}          
        &\makecell[l]{1.Time pressure leads to cognitive overloading;\\2.Increased demand for rapid reaction time;\\3.No time to consider the end-user;} \\
        \cline{3-4}
        \textbf{{Group 2}} 
        &
        &\makecell[l]{1.Unsatisfactory user experience;\\2.Stressful to receive voice response from system;\\3.Long waiting time for results;}
        &\makecell[l]{1.Perfectly satisfied with the revised strategy;\\2.Good collaboration flow and high quality outputs;\\3.Team felt more united than In previous trial;}\\
        \cline{3-4}
        \textbf{{Group 3}} 
        &
        &\makecell[l]{1.Lack of system feedback;\\2.The system was not providing\\ the required information;}
        &\makecell[l]{1.Dynamic workflow leads to enhance performance;\\2.Improved efficiency with speech recognition;}\\
        \cline{3-4}
        \textbf{{Group 4}} 
        &
        &\makecell[l]{1.Fulfills the basic need;\\2.Believe that there is no need for \\the note taking;}
        &\makecell[l]{1.New strategy is effective but rapid responses are hard to achieve;\\2.Inability to meet the requests or the end-user;}\\
        \hline
    \end{tabular}}
    \caption{\sy{The End-User experience of the first trial and the second trial after iteration, listed together with Wizards' perceptions about their task performance. }}
    \label{tab_strategyfeedback}
\end{table*}

\subsubsection{Wizard perceptions of collaboration effectiveness}
Wizards described the effectiveness of the collaboration in terms of how their work contributed to the team's goals and how the other Wizards supported them.

In \underline{the Pipeline Strategy}, one Wizard's work can aid subsequent tasks carried out by the other Wizards. Such support was perceived to reduce other Wizards' workload.
For instance in G1, the content labels produced by the Labelling Wizard were helpful to content separation, marking relevant key information and organising the summary. G1P1-HSW: ``After he [labeled] each piece of content, I could know what this piece of text was about [from] the beginning of each paragraph. I didn't have to read it again. Then, [we needed] to find some relevant content [and] keywords in this part.''
G4 indicated that highlighting and summarizing Wizard benefited from other Wizards rapid provision
of where start the work. As explained by G4P10-HSW: ``After they helped me sort [the text] out, I didn't have to read the questions part of the interview. After reading the paragraph with the label, I could select \sy{the relevant one.'' These Wizards'} work also contributes to the cooperation, for example, by helping the label Wizard to skim through the marked content and to add labels. G4P12-DW: ``He bolded the interview questions and then helped me to split the content. I then concentrated on the content of the interviewees' answers.''

With \underline{the Coordinated Strategy}, the Dictation Wizards in Group Two managed the rhythm of the dictation, allowing time for the Wizards to process the text. They also provided the groundwork and managed the workflow for the other two Wizards, which was well received. As G2P4-LW said``I was supported by [DW name] when I was labelling. She controlled the team rhythm and timing, which enabled me to have enough time to do the work more carefully. ''.   
In contrast to Groups One, Three, and Four, who optimized their workflow to focus on providing timely services to End-Users, the coordinating Dictation Wizard in Group Two interrupted the End-User to keep the pace for the Wizards' work. As they explained: ``If I hadn't interrupted the interview, [my fellow collaborators] wouldn't have had time to edit and label the content, which would have made it impossible to generate useful results (G2P5-DW)''. This also shows that, while all the Wizards shared the same goal, they had very different priorities in the moment.

Wizundry 2.0 better supported the division of work, with more clearly separated roles and interface components for Wizards. Nevertheless, work conflicts still occurred from time to time. \underline{The} \underline{Spontaneous Strategy} was prone to conflicts, especially when multiple Wizards edited the same line of text. As the Dictation Wizard in Group 3 described, 
``G3P7-DW: I wanted to change the word in that line or wrap it. Then the [LW name] side would start to delete words, and the [HSW name] side would start highlighting (...) When all the cursors appeared on the same line, since all of us were [working on the same area], everyone was anxious to complete their tasks, and everyone was focused on the latest line. It was easy to have problems.''
Even in the more coordinated Group Two, the Dictation Wizard mentioned being hesitant to take action when multiple Wizards were on the same line, which resulted in him delaying his response by trying to make room for the others. 
However, all groups were able to improve their coordination by adjusting their strategies based on the End-User's feedback. Most prioritized the feature explicitly requested by the End-User.

\subsubsection{Wizard Task Load.}

\begin{figure}[htbp]
  \centering
  \includegraphics[width=0.8\linewidth]{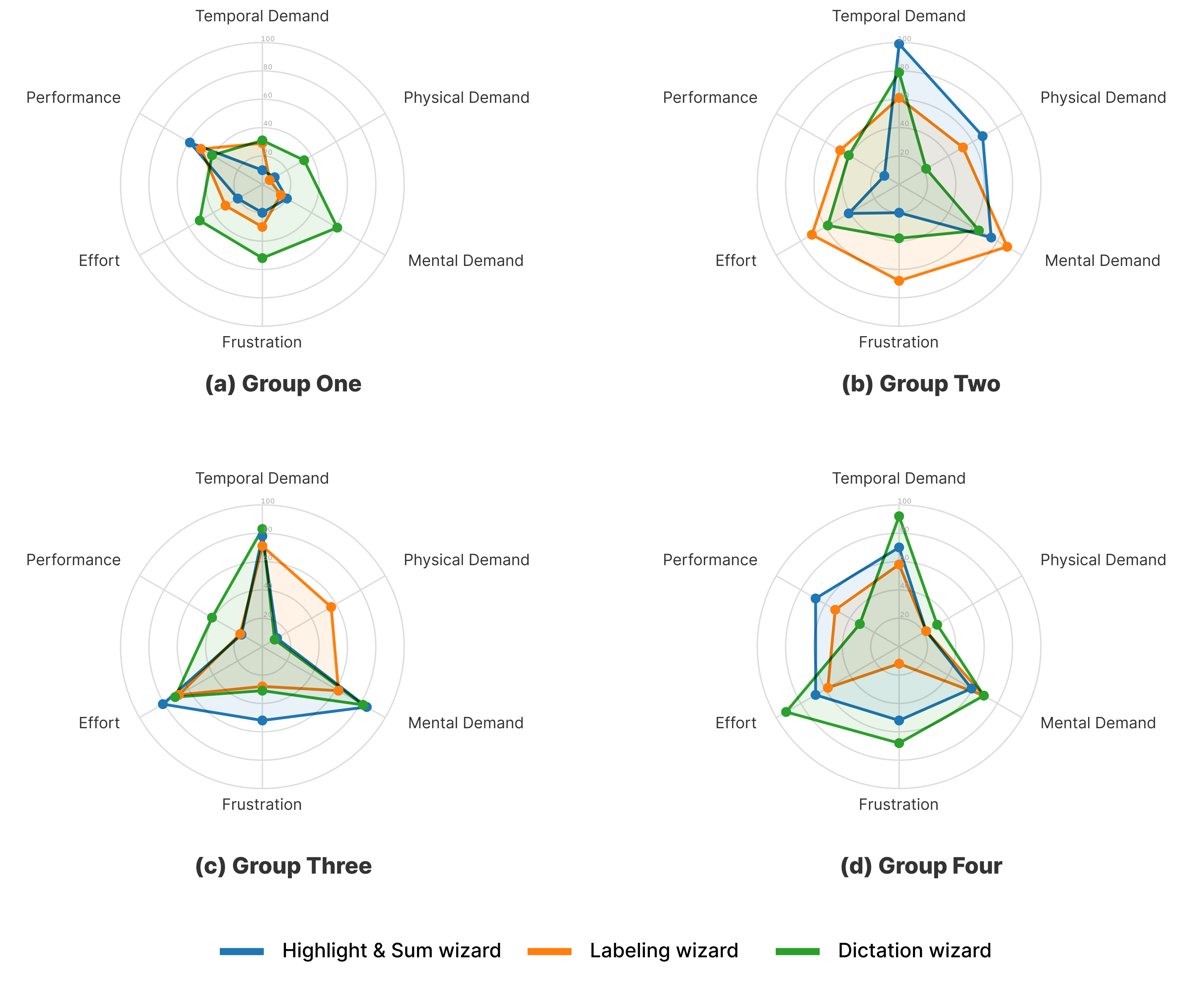}
  \caption{\sy{Average scores for the NASA-TLX on the Wizard side. Results for the three Wizards in each group. Each colour in the group represents different Wizard roles: Green Link means the Dictation Wizard, Orange Link means the Labeling Wizard, and Blue Link means the Highlight and Summary Wizard.}}
  \label{fig.Wizardnasarader-1}
\end{figure}

The raw, summed, and averaged NASA-TLX scores across multiple Wizards are presented in Appendix Table.\ref{tab:wizardnasa}).
Participants in the [Multiple-Wizard] condition had an overall average summed score of 29.4 (SD=7.4, MED=32, IQR=6.0) and mean score of 4.9 (SD=1.2, MED=5.3, IQR=1.0). 

We observed that the way the three Wizards cooperate with each other has an effect on their task load. As shown in Fig.\ref{fig.Wizardnasarader-1}, Group 2 (Coordinated Strategy) and Group 3 and 4 (Spontaneous and Pipeline Strategies) all had visibly higher overall task load than Group 1 (Pipeline Strategy). While differences in rating between groups exist, this still suggests that coordination overheads are high in our Multi-Wizard approach regardless of the coordination strategy.

In terms of the dimensions of task load, we can see that ``Temporal'', ``Effort'' and ``Mental'' load are the highest for all three high task load groups (G2, 3 and 4) (Fig.\ref{fig.Wizardnasarader-1}). Consistent with the NASA-TLX results, our participants expressed that the primary challenge of the Wizards' job was the time pressure. Wizards were required to make the shortest response time to provide a realistic experience for the end-user. Every group of Wizards mentioned this challenge, with one dictation Wizard explaining that ``It’s just too late to do it, they talked too fast, and there were two people, there were too many changes [that] needed to be edited, and [so there was just not enough] time. (G1P3-DW)'' G3P9-HSW said, ``I did not have time to listen to what was being said on the user side. Saw her results for highlighting. '' Similarly, the label Wizard said he could not use them all. Quoting G2P4-LW, ``Two speakers were easy to label, but there were, so many label types, I couldn't think of them in time.''


\subsubsection{End-User task load and performance.}


\begin{figure}[htbp]
  \centering
  \includegraphics[width=0.8\linewidth]{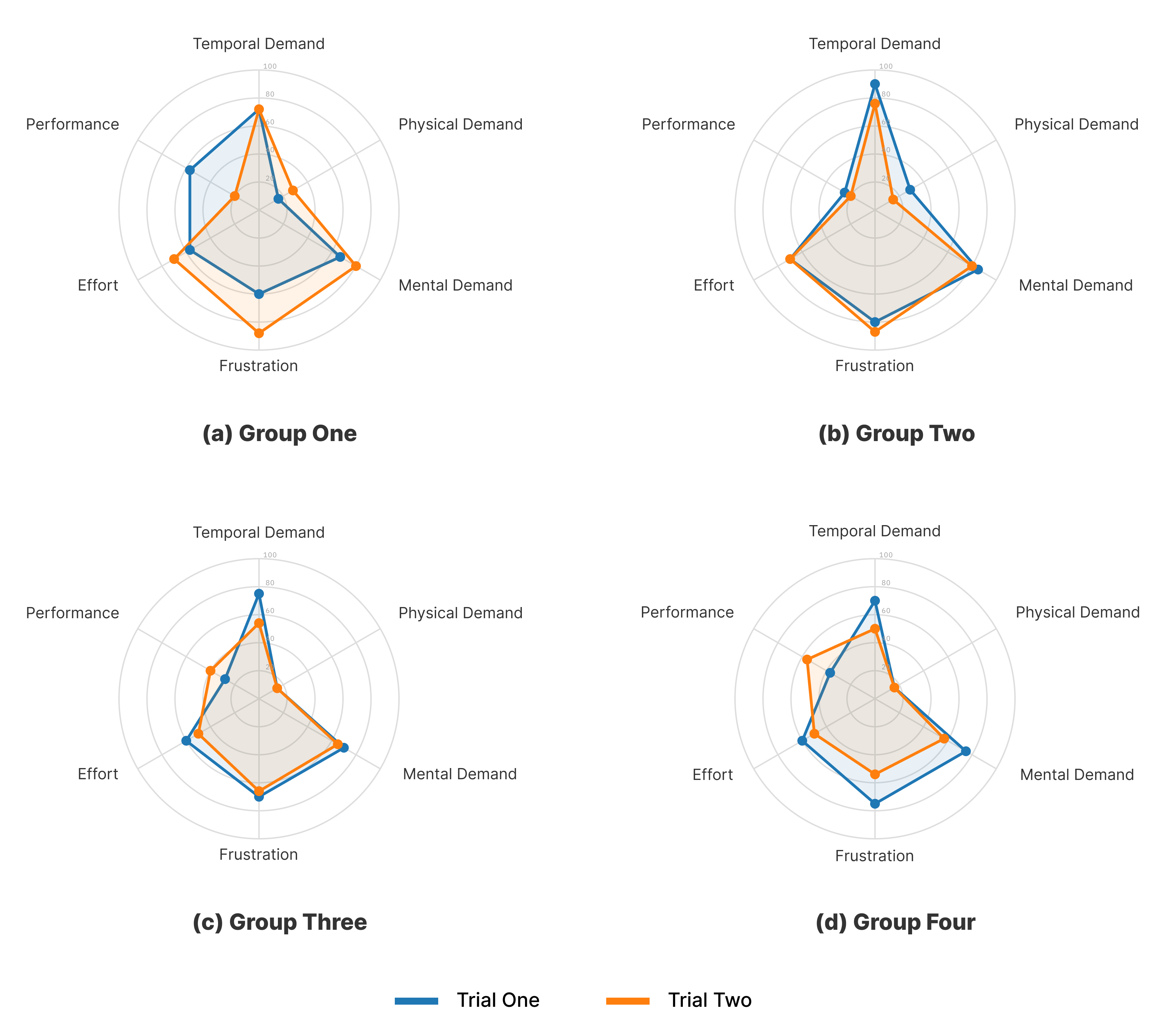}
  \caption{\sy{Average scores of End-User for each NASA TLX questionnaire; Results for two trials in each group; Each colour in the group represents a different trial; Blue Link means Trial 1, Orange Link means Trial 2.}}
  \label{fig.usernasarader}
\end{figure}

The raw, summed, and averaged NASA-TLX scores across End-users are presented in Appendix Table~\ref{tab:usernasatl}. 
The End-User who acted as the interviewer had an overall average summed score of 32.7 (SD=3.2, MED=32.8, IQR=4.8) and mean score of 5.5 (SD=0.5, MED=5.5, IQR=0.8). \sy{There are consistencies and differences in the End-User experience when different groups of participants performed as Wizards.} As per Fig.\ref{fig.usernasarader}, the overall ``shape'' of the task load across groups was similar, with ``Temporal demand'', ``Frustration'', ``Mental demand'', and ``Effort'' scoring relatively high. This reflects the nature of the task and the fairly consistent execution of this task by the confederates on the End-User side. Comparing Trial 1 and Trial 2 for each group, we can get a sense of whether and how each provided an improved End-User experience after reflecting on and updating their strategies. Most groups (Groups Two, Three, and Four) managed to reduce the End-User task load to various degrees, while Groups Three and Four improved the perceived task performance of the End-User. One potential reason is that these two groups optimized for providing what the End-User requested after Trial 1. Group One, as the exception, provided a poorer End-User experience after the iteration, which led to higher Effort, Frustration, Mental Demand, and lower Performance scores. This was the only group that solely used the Pipeline Strategy in the second Trial. While this strategy made it easier for the Wizards to work together, it caused longer waiting times for the End-User, and thus negatively impacted the End-User experience.

\subsubsection{Understanding the gaps between Wizards and End-Users}

\paragraph{Lack of feedback from the simulated system.}
The main challenge mentioned by End-Users when interacting with the simulated system was the lack of system feedback. 
As G2U1 explained: ``I did not know whether the system works or not, and I wasn't sure how to use it properly.'' When users found the system unable to provide feedback signals, we observed that they tried to interact with it through voice commands but we did not respond. 

On the Wizard side, we observed that only the Dictation Wizard in Group Two used the ``speech boxes'' to communicate with the End-User. However, their purpose was to orient the End-User to the work rhythm of their team. As G2P2-DW said: ``[HSW name] told me to tell the user to pause and wait for them, or it would not be done.'' The Dictation Wizard in Groups One, Two, and Three kept the ``Microphone trigger'' open and did not use the ``speech boxes''. The rest of the Wizards indicated that the individual tasks were highly demanding and so they did not have any extra time to allocate to End-User interaction. For example, as G3P9-HSW said: ``I did not have time to ask the user, and I did not have time to ask him to stop the interview.'' 

\paragraph{Wizards could not satisfy End-User requests.}
End-Users complained that the system did not meet their basic expectations for assisting them with the interviewing task. 
Even though End-Users provided specific requirements on the assistance that they needed from the system, including the labels and type of content to highlight, the simulated system did not meet their requirements.
As G3U1 said: ``I wanted to distinguish between questions and answers [in the transcription] by  bolding the questions, but this simple function was not achieved. '' 

On the Wizard's side, participants explained that it was difficult to distinguish the points that needed to be highlighted when under time pressure. They did not have sufficient time to react and handle all of the tasks in a time-sensitive fashion while matching the pace of End-User's speech. One End-User attempted to give hints to the system, i.e., the Wizards, through speech: ``The second time I said that I wanted to emphasize some of the content in summary. so I said `Yes, it is a key point' or `These need to be written down' (G3U1).'' However, such requests were not carried out by the Wizards. 
One Wizard explained why some requests were not picked up, G3-HSW explained: ``I heard it but I didn't have time to do it. Sometimes I was still writing the summaries.''

\paragraph{Need for better ways to communicate End-User needs to Wizards.}
The End-Users expressed the lack of methods to correct the system when the system tagged or highlighted content wrongly.
G4U1 mentioned: ``When I was conducting an interview, I saw the recorded content, and when I scanned the highlights, I found it was not the focus of what I wanted to record. I told the system which content is very important, and it didn't work. I also could not edit or tag these anytime.'' 

On the Wizard side, three groups (G1/3/4) mentioned that they wished to know the agenda and questions of the interviews from End-Users in advance. A Dictation Wizard commented, ``It could save lots of time in dictation and editing (G4P12-DW).'' A similar expectation was expressed by another Wizard, 
who wished to know the agenda and topic in advance, which would help predict the content of the task in the moment: ``At least, we used to able to prepare what we were going to hear, and when we saw the interview questions, we would probably predict the content of the dictation first, which might be much more efficient (G3P9-HSW).''

\paragraph{Mismatched perceptions of task performance.}
The perception of how the task went could vary greatly between the Wizards and the End-User in the same trial. As per Table~\ref{tab_strategyfeedback}, Group Two thought that the Coordinated Strategy they chose was appropriate and useful. However, the experience of the End-User was hard to balance: End-Users disliked being interrupted verbally by the system
and the long waiting times for system responses. Other groups' perceptions were not far off, yet seemed to be much more affected by their collaboration experiences with teammates rather than thinking from the End-Users' perspective.

\subsubsection{Feedback for the design of Wizundry 2.0}

\sy{Twelve Wizard participants expressed that the features and design of the system were simple, user-friendly (Wizard-friendly, to be clear), and effective in helping them perform their tasks. Notably, the cursors helped them reduce workplace conflicts and be aware of the other Wizard's work. As one Dictation Wizard explained, ``G3P7-DW: That real-time cursor, I would not move my cursor after seeing it. I knew they were changing this, too. When they finished, I'd see if I needed to continue to do work there or not. I wouldn't go near where the cursor was until it was gone.'' 
Moreover, participants appreciated the division of labour supported by the interface. ``[This] helped to reduce the burden on my side. I could focus on my part and not worry about theirs (G1P3-DW).''
Participants also appreciated the modular interface, which allowed them to add or remove functions and components to customize their workflow. For instance, some Wizards made great use of the speech boxes. 
As G3P7-DW noted, ``[HSW name]  told me about some requirement, and I added some responses. (...) Then I used each new response. For my partners, they feel a bit better.'' }

\section{DISCUSSION AND FUTURE WORK}
In this section we first discuss our findings on the feasibility, effectiveness and challenges of the Multi-Wizard approach as well as its design opportunities. Then we elaborate on the generalizability of our approach and findings, as well as the trade-offs we had to choose and the limitations of this work.

\subsection{Feasibility and effectiveness of the Multi-Wizard approach}
Both studies 1.0 and 2.0 showed that having Multiple Wizards sharing the workload of simulating a real-time intelligent system is a feasible approach, sometimes necessary. Can Multiple Wizards provide a better experience than a single Wizard? Our studies indicated that this is certainly possible. In study 1.0 we tested with a relatively simple system that handles only text input and editing. The task distribution between Wizards were mostly uneven (Fig 3), where one Wizard did most of the work and the other was assisting. An interesting effect we observed was that not only the \emph{action} of other Wizards, but also the \emph{anticipation of action} by other Wizards brought much stress to the team in fear of conflicts. Since being a Wizard for a real-time interactive system is such a time-critical task, the uncertainty that emerge from overlapping task areas amongst participants was highly detrimental because participants had to pay extra attention to avoiding conflicting edits. Due to this cost and the coordination overload, we received mixed feedback from Wizard participants about whether it was actually helpful to have one more Wizard doing the task. Therefore we designed study 2.0 to test a scenario where clearer division of labor can be made and where the Wizards needed to simulate more advanced system features. The result was positive for study 2.0 as Wizards found the teammates' work helpful and essential, and the system was able to support their collaboration effectively. From 1.0 to 2.0 our findings show that a Multi-Wizard approach works better for simulating complex systems with clearer separation of responsibilities among Wizards. 

Besides sharing workload, we found that Wizundry is an effective platform for collecting and testing creative ideas for designing intelligent system behaviours. The Wizards in our studies came up with novel collaborative strategies and system behaviors. In study 1.0, the Wizard participants came up with interesting ways of dividing their tasks: content based, interface-controller based and input-device based approaches. Some dominant collaboration styles also emerged to reduce conflicts. \newsy{Under the dyadic approach, Wizards focused on how to divide work and complete the same task. Wizundry was designed to allow Wizards to self-organize and divvy up tasks. We discovered that some participants actively support their partner, which was usually disruptive.} In study 2.0, the Wizards redesigned their workflow by making sub-tasks sequential, or even by using the speech boxes to negotiate the pace of work with the end-user. These findings were a surprise to us and made us realize an unexpected use of Wizundry platform as an ideation + evaluation tool for designing intelligent system features with very fast iteration capability. 

\newsy{The Dyads and Triads studies showed the possibility for facilitating cognitively demanding tasks for wizards. Wizundry presented a novel Multi-Wizarding platform to participants acting as Wizards, assisting tasks that were cognitively demanding for them. Wizards were tasked with simulating advanced functions and features while dynamically responding to unpredictable requests, in a speech-based AI interaction context. As such systems become more innovative, users will continue to expect more advanced functionality and open-ended engagement from Wizards. The ability of each Wizard to grapple with the demands of such advanced tasks and system complexity will affect the degree to which the simulation is perceived as realistic and believable. Essentially, this is a matter of the cognitive demands placed on each Wizard. For us, this finding meant designing Wizundry to address these challenges in appropriate and useful ways. Including multiple Wizards to offload cognitive and task labor was but one design strategy. The used approach to the design of Wizundry appears to have helped facilitate Wizards in handling cognitive demands across different tasks.}

\subsection{Challenges and opportunities in Multi-Wizarding}
Although previous works explored the challenges of being a Wizard \cite{shin2019apprentice}, we are not aware of any existing work of formal studies evaluating Wizard experiences. Our findings revealed empirical insights on the challenges of Wizarding and their collaboration, leading to questions for future research.

The primary challenge we faced was the coordination between Multiple Wizards. 
Although we improved on reducing potential conflicts from version 1.0 to 2.0 by conducting our study in a shared physical space where Wizards can interact more fully and in real-time, it still happened from time to time. In both studies there was a voluntary emergence of leaders in order to help coordinate the actions of the team and even to interact with the end-user to negotiate the task rhythm. Such collaboration style was mostly appreciated by other team members. However, all the Wizards in both studies reflected the high temporal pressure of the task, which led to little, or no time, for verbal communication between Wizards. 
\cltext{Work conflicts occurred amongst multiple Wizards in our studies, especially when the responsibilities were not clearly separated. These findings are inline with previous works in terms of coordination and synchronization, we also found challenges in our experiments, especially with editing conflicts and workload management, which are similar to  other real-time collaborative work contexts involving multiple people (e.g., e-work and online teaching) \cite{knister1990distedit, nof2003design}. The collaborative cursor provided in our system was a basic function to provide awareness of Wizards' actions among them. Moving forward, much more could be done to further improve this.} 

\cltext{Previous studies on computer-supported working mechanisms addressed these challenges by enhancing the exchange of information between team members, such as updating decision-making algorithms and establishing alternative communication channels \cite{shamekhi2018face}. Another way was to design advanced user interfaces and use rhetorical visualisation to enhance users' interaction effectively, which may help reduce the cognitive load \cite{huang2017supporting}. Some studies focused on using designed workflow to handle and segment complex tasks, especially in the crowd-sourcing field \cite{knister1990distedit, kulkarni2012collaboratively}. To address editing conflicts in the co-writing system, some solutions aim to track the working actions and present the area of issues in the system interface, allowing the user to resolve it themselves \cite{knister1990distedit}. Such solutions could be used in future development of Wizundry systems to better assist Wizards coordination. 
Moreover, we can draw further inspiration from literatures in ``Social translucence'', a theory-based design approach proposed by Erickson and Kellogg in 2000 \cite{erickson2000social}. They proposed a framework for designing coordination mechanisms and collaboration norms by deploying a shared visualization of each user's activities. 
Following this line of thinking, we can imagine that future versions of Wizundry could visualize other signals of Wizards' activities, such as their gaze focus locale, to provide more timely collaborative awareness.
}

The second challenge we identified in both studies was the synchronization and communication between Wizards and the end-user. The end-user's locale of attention was not obvious for the Wizards. In a dictation and text processing scenario, this was particularly problematic as the lengthy dictation led to Wizards losing track of where the End-User or other Wizards are.
We believe this is an interesting design challenge that can be tackled in future work, to enhance communication between users and Wizards or intelligent systems. Moreover, our findings in the end-user experience of study 2.0 showed the importance and promising effects of providing customized assistance as requested by the end-user. It also appeared to us that it may be better to give part of interface control to end-users, to support close collaboration between the user and the Wizards, rather than simply letting the user verbally tell the system what she/he wanted. We will explore this in our future work. Last but not the least, findings in study 2.0 showed that prior knowledge of the end-user's agenda, goals and preferences were much needed for Wizards to assist them effectively. Thus a prior ``system configuration'' with end-users' input is a necessary step for future studies alike, since what the Wizards felt would be a good experience may be completely off for end-users (example Group 2 in Study 2.0). Human-simulated systems carry human biases determined by their personal experiences and backgrounds. Such biases can be more than rule-based or data-trained computer systems, and vary a lot across different Wizards.

\subsection{From collocated to remote Multi-Wizards}
\cltext{
The Wizards in our experiments occupied the same physical room to complete their respective tasks. Since they were in the same space, they could freely communicate face-to-face and use each other's screens to negotiate their work during the experiment. 
In the discussion sessions where the Wizards were making strategies, we observed Wizards discussing with each other by pointing each other's screen and using hand gestures to facilitate explanations. These verbal and gestural communication was effective and efficient in supporting their discussion. However, we rarely observed such communication when they were performing the tasks. This was due to the high temporal pressure when trying to follow the speech input of End-Users - they did not have time to communicate. Therefore, we believe Wizundry could be used with Wizards located remotely, as long as they plan ahead before tasks starts with a video conferencing call while sharing screens.}

\subsection{Generalizability}
While our case studies focus on smart dictation interfaces with specific features, the findings can be generalized into other Speech-to-Text applications and development of human-AI interaction.
First, our findings provide insights and design inspirations for the development of Speech-to-Text applications such as the voice typing interfaces embedded in modern smartphones, AI-powered transcription tools like Otter.ai, smart note-taking tools for meetings, conferences, etc. The dictation and editing tasks we tested are fundamental for speech-based text input interfaces, and the intelligent text processing features (keyword spotting, text categorization and summarization) we provided in study 2 are being developed in the field of NLP and Knowledge Graph for years \cite{yoo2020intelligent,haase2017alexa}. As these technologies get increasingly powerful, they will be used more and more to power real-time interfaces. 

Second, although our findings in particular task strategies and system features are specific for dictation-based tasks, the higher-level findings can be generalized for the design of other human-AI interactions. 
For instance, our findings suggest that other researchers could implement a Multi-Wizard platform for other research contexts, e.g., multimodal interfaces \cltext{or human-robot interaction}, and use it as an ideation and fast iteration tool for designing intelligent system features. 

\cltext{Our identified cooperation strategies between Wizards and their effects on End-User experiences are generalizable to different contexts of use. We found: a sequential Pipeline strategy was easier for the Wizards but hindered End-User experiences; coordination overhead was heavy regardless of whether it was spontaneous or managed by a leader. Future work shall be aware of this trade-off when designing their own platform.} 

What we learned about clear task distribution in view of temporal pressure highlight potential challenges in other contexts too. 
\sy{Work conflict would also appear in other contexts. The importance of providing transparency of each Wizard's actions shall be universal. A modular system did provide the flexibility and support to divide tasks and customize interfaces for individual Wizard, therefore shall be effective in other contexts as well.} 

Last but not least, our findings about the mismatched perceptions on task performance and the need for Wizards and End-User communication and collaboration were also generalizable to other contexts. 
It was interesting to see the Wizards' perception of their task performance could be the opposite to what the End-User experiences. This suggests a need for ways to direct Wizards as to set their priorities better aligned to the ones of End-Users. We also found it was important to provide the possibility for End-Users to correct Wizards' work and give real-time requests. In addition, we found that prior knowledge about the agenda of the end-users could be helpful for Wizards to prepare for the task and pay attention to the important information. 


\subsection{Methodology for evaluating a multi-wizard system}
Wizundry is a platform supporting two user groups-- Wizards and end-users. To have certain control over confounding factors, we employed researchers as confederates posing as end-users for both studies. The confederates were given pre-formulated scripts to read (Study 1.0) or a topic guide for the interview task (Study 2.0). Now that we have gained understanding on the Wizards' side of the system by controlling the end-user side, our future work will conduct studies with real participants as end-users.

Our study design from 1.0 to 2.0 were iteratively adjusted. In Study 1.0 we let the Wizards decide their collaborative strategy as in how they divided responsibilities and controls. The observations from this study made us realize a clearer division of roles was beneficial or even necessary for a multi-wizard system due to the temporal demand of the task. Therefore in Study 2.0 we enforced a role division by default, which allowed us to dive deeper into the understanding of how workflow can be managed and their impact on end-user experiences. This 2-step methodology may be used in designing future studies for other cooperative systems like Wizundry.

At last, we shall be aware of the limitations of our approach. Our own researchers are not eligible to evaluate the system as real end-users, thus their reported experiences were not taken as objective evaluation of our interface. However, collecting the end-users' experience was important for our study. The purpose of having them report their experiences entailed two objectives: 1) to provide end-user feedback to Wizards to test whether they can effectively adapt their strategy and workflow to achieve a better end-user experience; 2) to compare the Wizards' collective performance across groups, as the researcher's subjective biases would be consistent across different Wizard groups.

\subsection{\newsy{Trade-offs between labor division strategies in Multi-Wizarding}}

\newsy{There were tradeoffs for dividing labor: start with clearer divisions in strategies versus without any plans. Wizards who started without a plan tended to have incorrect assumptions regarding the division of labour. Even though everyone’s initial intention was to enhance productivity, the resulting unclear divisions of labor were averse to the group's ability to work effectively. The efficiency of completing tasks and sharing the workload were considerations for dividing labor. On the other hand, starting with clear division would enable each wizard to define his or her tasks and adhere to the specified execution plan. However, excessive focus on individual tasks may result in a lack of communication and coordination.}

\newsy{Multi-Wizard approaches, regardless of whether there is a defined strategy for assigning tasks to different Wizards, offer significant potential for WoZ contexts, especially involving voice and speech. Wizundry allows for combining modular features (e.g., Feature Selection List in Landing Interface) with independent Wizard work. Such Multi-Wizard configurations can support a single Wizard at the start of a WoZ study, which can then grow into a dyadic or Multi-Wizard setup as needed, based on the division of labor. By evaluating diverse groups of Dyads and Triads, we found that the challenges and the trade-offs were different for each labor allocation method that individual groups came up with and  optimization. Researchers can explore the right combination of attendance, system features, and best-performing cooperation strategies with a greater diversity of Wizards and teams. Further, the collaborative space of each Wizard in each context may not be static. We observed that dyads began spontaneously changing strategies when they realized that their task performance was unsatisfactory. This may have been affected by the physical environment of the experiment, including device placement and personal operating preferences. We did not record the individual habits of each Wizard, which future work can consider.}

\subsection{\newsy{Advancing WoZ methodology}}

\newsy{The Multi-Wizard approach enables co-creation and co-learning amongst independent wizards in order to successfully complete the WoZ experiment. This approach provides the opportunity to brainstorm ideas, avoid self-centered perspectives, and expand creativity by drawing on the insights of others. Combining these diverse ideas of wizard members helps create more effective solutions to demanding tasks. By leveraging multiple Wizards’ knowledge, experience and brain power, we have the opportunity to rapidly prototype more powerful systems, such as more intelligent features, multi-modality input, etc. However, this is not the solution to address all the problems. The creativity and openness of advanced forms of human-AI interaction presents new challenges for researchers who wish to conduct WoZ studies. Having Wizards simulate a system still has its inherent limitations, including the relatively long reaction and computation time humans need, and the cost of communication and coordination between multiple of them. Future WoZ methods need to address these challenges through other approaches.}
\section{Conclusion and future work}
We have presented a research toolkit called Wizundry and the iterative design and evaluation process we undertook through versions 1.0 to 2.0. To our knowledge, this is the first effort to systematically explore the possibility of leveraging multiple Wizards to expand the capability of researchers in simulating intelligent systems that are hard to simulate by one Wizard \newsy{due to high cognitive demand}. It is also the first work that employs designs inspired by real-time crowd-sourcing platforms in a WoZ research context. Our studies revealed the feasibility and effectiveness of the Multi-Wizard approach, garnered insights on the challenges and opportunities involved, and inspire directions for future research. Our toolkit provides a modular interface built on an extensible software framework, and will be made available in an open source format for the research community. Our findings showed Wizundry is an effective WoZ platform as it could help researchers simulate and test intelligent features for a future dictation system, as a start point. By conducting studies with a simulated system supported by Wizundry, researchers could identify the important factors that improve or hinder end-user experiences. Different end-user experiences can be provided by adjusting priorities for the collective goal of the Wizards. We also found insights about how different cooperation strategies and division of labour could affect end-user experiences. Our work serves as an example of using a Multi-Wizard system to ideate and quickly iterate design concepts for human-AI interactions.  

\subsection{\cltext{Future work}}
Wizundry presented in this paper is our first attempt to systematically explore possible ways to improve the methodology of WoZ studies by supporting Multiple Wizards in close cooperation. We learned important lessons through the major and minor iterations from Wizundry 1.0 to Wizundry 2.0. While this is by far not the end of the exploration, this work lays the ground by providing a web-based toolkit that is modular and extendable, for future studies in the research community. Based on the feedback we received from our studies, the next version of Wizundry will seek new ways of reducing temporal pressure for Wizards. It appeared to us that the temporal pressure of being a Wizard cannot be alleviated by adding more Wizards. Other approaches will be explored, including better visualization of the end-user and Wizards actions. We will also explore other ways of providing feedback to end-users beyond providing speech boxes.

\begin{acks}
This research was supported by the Hong Kong Research Grants Council - ECS scheme under the project number CityU 21209419. 
\end{acks}

\bibliographystyle{ACM-Reference-Format}
\bibliography{multi-wizard}

\clearpage
\appendix
\section{Appendix - NASA-TLX scores}


\begin{table*}
  \small

  \begin{tabular}{ccccccccccc}
    \toprule
    ID & Mental & Physical & Temporal & Perform. & Effort & Frust. & Sums & Means & St. Dev. & Inter. Q.\\
    \midrule
    1 & 9 & 7 & 8 & 5 & 8 & 8 & 45 & 7.5 & 2.5 & 4\\
    2&7&3&2&6&7&1&26&4.3&2.1&1.8\\
    3&10&1&8&4&10&8&41&6.8&2.2&2.3\\
    4&7&2&4&4&7&4&28&4.7&2.1&4.8\\
    5&7&8&3&9&6&3&36&6&1.4&2\\
    6&4&3&4&7&6&2&26&4.3&0.4&1\\
    7&5&5&5&5&6&5&31&5.2&2.6&4.5\\
    8&5&6&6&4&5&5&31&5.2&1.8&2.8\\
    9&4&4&5&7&3&1&24&4&1&1.8\\
    10&7&6&7&7&7&6&40&6.7&2.5&3.8\\
    11&5&3&6&9&5&3&31&5.2&1&1.5\\
    12&6&3&5&1&5&4&24&4&2.7&6.8\\
    \bottomrule
  \end{tabular}
   \caption{NASA-TLX scores in Prototype 1.0 for Wizards.}
  \label{tab:nasatlx}
\end{table*}

\begin{table*}
\small
  \begin{tabular}{ccccccccccc}
    \toprule
    ID & Mental & Physical & Temporal & Perform. & Effort & Frust. & Sums & Means & St. Dev. & Inter. Q.\\
    \midrule
    1& 2&1&1&6&2&2&14&1.3&1.7&1.0\\
    2& 2&1&3&5&3&3&17&2.7&1.5&1.5\\
    3& 6&3&3&4&5&5&27&4.6&1.2&2.0\\
    4& 9&2&6&5&7&6&36&6.0&2.2&2.4\\
    5& 6&5&8&4&6&4&33&5.5&1.5&2.2\\
    6& 8&7&10&1&4&2&32&5.3&3.3&5.5\\
    7& 8&1&8&4&7&3&33&5.4&3.0&5.1\\
    8& 9&1&8&2&8&5&33&5.5&3.3&6.6\\
    9& 6&6&7&2&7&3&30&5.1&2.3&4.1\\
    10& 7&3&9&3&9&7&39&6.5&2.7&6.0\\
    11& 6&2&7&7&7&5&34&5.6&1.8&1.6\\
    12& 7&2&6&5&6&1&27&4.5&2.2&3.6\\
    \bottomrule
  \end{tabular}
   \caption{NASA-TLX scores in Prototype 2.0 for Wizards.}
  \label{tab:wizardnasa}
\end{table*}


\begin{table*}
    \small
  \begin{tabular}{ccccccccccc}
    \toprule
    ID & Mental & Physical & Temporal & Perform. & Effort & Frust. & Sums & Means & St. Dev. & Inter. Q.\\
    \midrule
    G1T1 &    7&2&8&6&6&6&35&5.4&2.1&1.0 \\
    G1T2 &    8&3&8&2&7&9&37&5.9&2.9&5.5\\
    G2T1 &    9&3&9&3&7&8&39&6.3&3.0&6.0\\
    G2T2 &    8&2&8&2&7&9&36&5.8&3.1&6.0\\
    G3T1 &    7&2&8&3&6&7&33&5.3&2.6&4.5\\
    G3T2 &    7&2&6&4&5&7&31&4.8&1.9&2.5\\
    G4T1 &    8&2&7&4&6&8&36&5.5&2.5&4.0\\
    G4T2 &s   6&2&5&6&5&6&30&4.7&1.6&5.0\\
    \bottomrule
  \end{tabular}
  \caption{NASA-TLX scores in Prototype 2.0 for End-users.}
  \label{tab:usernasatl}
\end{table*}

\received{January 2022}
\received[revised]{July 2022}
\received[accepted]{November 2022}
\received[published]{April 2023}

\end{document}